\documentclass[twocolumn]{aastex61}

\usepackage{verbatim}
\usepackage{apjfonts}
\usepackage{ulem}
\usepackage{mathptmx}
\usepackage{natbib}
\usepackage{soul}
\usepackage{graphicx}
\usepackage{xcolor}

\newcommand{\chandra}{ {\it Chandra} }

\newcommand{\spitzer}{ {\it Spitzer} }
\newcommand{\akari}{ {\it AKARI} }
\newcommand{\ciber}{ {\it CIBER} }
\newcommand{\mum}{$\,\mu$m}

\newcommand{\new}{{}}

\begin{document} 

\title{The SPLASH and~$\chandra$~COSMOS Legacy~Survey: the cross power between near-infrared and X-ray background fluctuations}

\author{Yanxia Li}
\affiliation{Institute for Astronomy, University of Hawaii at Manoa, Honolulu, HI 96822, USA}
\author[0000-0002-1697-186X]{Nico Cappelluti}
\affiliation{Physics Department, University of Miami, Coral Gables, FL 33124, USA}
\author[0000-0001-8403-8548]{Richard G. Arendt}
\affiliation{Observational Cosmology Laboratory, Code 665, Goddard Space Flight Center, 8800 Greenbelt Road, Greenbelt, MD 20771, USA}
\affiliation{CRESST II / University of Maryland, Baltimore County, 1000 Hilltop Circle, Baltimore, MD 21250, USA}
\author[0000-0002-0797-0646]{G\"{u}nther Hasinger}
\affiliation{Institute for Astronomy, University of Hawaii at Manoa, Honolulu, HI 96822, USA}
\affiliation{European Space Astronomy Centre (ESA/ESAC), Director of Science, E-28691 Villanueva de la Ca\~nada, Madrid, Spain} 
\author{Alexander Kashlinsky}
\affiliation{Observational Cosmology Laboratory, Code 665, Goddard Space Flight Center, 8800 Greenbelt Road, Greenbelt, MD 20771, USA}
\affiliation{SSAI, Lanham, MD 20706, USA}
\author[0000-0002-4326-9144]{Kari Helgason}
\affiliation{Centre for Astrophysics and Cosmology, Science Institute, University of Iceland, Dunhagi 5, 107 Reykjavík, Iceland}



\begin{abstract}

We study the source-subtracted near-infrared and X-ray background fluctuations of the COSMOS field using data from the~$\spitzer$~SPLASH program ($\sim$1272 hours) and\chandra COSMOS Legacy Survey (4.6 Ms). The new auto power spectra of the cosmic infrared and X-ray background fluctuations reach maximum angular scales of $\sim$ 3000$''$ and $\sim$ 5000$''$, respectively. We measure the cross power spectra between each infrared and X-ray band and calculate the mean power above 20$''$. We find that the soft X-ray band is correlated with 3.6 and 4.5\mum\ at $\sim$ 4 $\sigma$ significance level. The significance between hard X-ray and the 3.6\mum~(4.5\mum) band is $\sim$~2.2 $\sigma$ ($\sim$~3.8~$\sigma$). The combined infrared (3.6 + 4.5\mum) data are correlated with the X-ray data in soft ([0.5-2] keV), hard ([2-7] keV) and broad ([0.5-7] keV) bands at $\sim$~5.6~$\sigma$, $\sim$~4.4~$\sigma$ and $\sim$~6.6~$\sigma$ level, respectively. We compare the new measurements with existing models for the contributions from known populations at $z$$<$7, which are not subtracted. The model predictions are consistent with the measurements but we cannot rule out contributions from other components, such as Direct Collapse Black Holes (DCBH). However, the stacked cross-power spectra, combining other available data, show excess fluctuations about an order of magnitude on average at $\sim$4$\sigma$ confidence at scales within $\sim$300$''$. By studying the X-ray SED of the cross-power signal, assuming no significant variation from the infrared, we find that its shape is consistent with DCBHs.

\end{abstract}

\keywords{cosmology: observations --- infrared: diffuse background --- quasars: supermassive black holes --- X-rays: diffuse background}



\section{Introduction}\label{s:intro}

The Cosmic Infrared Background (CIB), the integrated emission from cosmic sources including those that are too faint to be resolved using contemporary observing techniques, establishes a unique measure of the unresolved sources in the Universe \citep{Hauser01,Kashlinsky05a}. 

The photons from sources in the reionization era are redshifted and their emission is expected to peak at the near infrared (IR) wavelengths at present time. Therefore, measuring the CIB serves as an important indirect method to study the early stars and black holes (BHs) which could contribute to the background emission \citep{Partridge67a,Partridge67b,McDowell86,Bond86}. 

However, measuring the absolute CIB is difficult due to the foreground contamination especially by our Galaxy and solar system, e.g., zodiacal light.  Measuring CIB fluctuations is a promising alternative, because it is much less sensitive to the absolute zero-point of the measurements and the zodiacal light is very smooth in the fluctuation spectra \citep{Kashlinsky96}. The fluctuation analysis can efficiently distinguish early populations, if any, from local stars and galaxies due to the distinct spectral amplitude and structure of the high-$z$ emission. Although the early populations are not individually resolved, their fluctuations could be amplified due to the clustering properties of the early populations \citep{Cooray04, Kashlinsky04}, and is therefore detectable via the CIB fluctuations after removing resolved objects in the foreground down to certain faintness levels. 

In fact there have been a series studies of the CIB fluctuations using various instruments. For example, using\spitzer IRAC, \citet{Kashlinsky07c, Kashlinsky07b, Kashlinsky07a, Kashlinsky12} reported consistent measurements of the background fluctuations in various fields up to $\sim$1000$''$, e.g., the Ultra Deep Survey \citep[UDS, ][]{Fazio11} and the Extended Groth Strip \citep[EGS,][]{Davis07,Goulding12}. After removing the resolved sources down to m$_{AB}$$\sim$25 mag, they found excess CIB fluctuations that are significantly higher than those from known galaxies on larger angular scales and have a different spatial distribution. 

Several scenarios have been proposed to explain the origin of the excess CIB fluctuation, including early populations (e.g., stars, BHs) and low-$z$ sources. Using the Japanese infrared astronomical satellite $\akari$, \citet{Matsumoto11,Seo15} suggested that the source-subtracted CIB fluctuations at 2.4, 3.2 and 4.1\mum\ around scales of a few 100$''$ may be caused by high-$z$ sources. Moreover, the SED of the fluctuations has a near-infrared slope of $\nu I_\nu \sim \lambda^{-3}$ out to 2.4 \mum, similar to the Rayleigh-Jeans tail of a blackbody distribution. The Lyman-breaks for those high-$z$ sources might be located at shorter wavelengths $\lesssim$ 2.4\mum, indicating an upper limit of $z$$<$24. 

On the other hand, \citet{Cooray12a, Cooray12} carried out their own studies of CIB with\spitzer and found consistent measurements but instead interpreted the CIB origin from the intrahalo light (IHL, tidally stripped stars during galaxy mergers and collisions) at redshifts of $\sim$ $z$ = 1-4. This scenario was also supported by \citet{Zemcov14} based on the $\ciber$ observations. The $\ciber$ CIB results, however, appear at variance with the earlier and deeper CIB fluctuation measurements reported from 2MASS \citep{Kashlinsky02,Odenwald03} and NICMOS studies \citep{Thompson07a,Thompson07b}.
 
Multi-wavelength studies can provide a more comprehensive understanding of the signal and differentiate between interpretations of the origin. As the first cosmic background emission that has been discovered \citep{Giacconi62}, the cosmic X-ray background (CXB) is mainly composed of AGN emission at redshifts up to 5 \citep{Boldt87,Brandt05,Hasinger08,Brandt15}. However, there might also be significant contributions from unresolved source populations, such as heavily shrouded Compton-thick AGN, which do not violate the observational constraints in the mid IR \citep{Comastri15}. A cross correlation between CIB and CXB not only provides a more useful tool to uncover where the excess CIB fluctuation arise from but also enables a better understand of the unresolved CXB contributors (see review by \citet{Kashlinsky18} for summary and discussion).

There have been a few measurements of the CIB and the CXB cross power and coherence \citep[][C17]{Cappelluti13,Mitchell-Wynne16,Cappelluti17c}. For example, \citet{Cappelluti13} used the~$\chandra$~deep images and found a 3.5-5~$\sigma$ correlation between the CIB, from~$\spitzer$/IRAC data (3.6$\mu$m and 4.5$\mu$m), and the unresolved CXB at soft band, on scales around $\sim$ 600 - 1000$''$. The cross-power was interpreted as imprints from early accreting black holes, e.g. by Direct Collapse Black Holes \citep[DCBH,][] {Yue13,Yue16} at $z$$>$12 or Primordial LIGO-type Black Holes \citep[PBH,][] {Kashlinsky16}. 

However, these studies of the CIB-CXB cross power are limited to an angular scale within $\sim$1000$''$ due to the small fields studied, although a larger field is better, given that the power spectra at the larger angular scale are less dominated by the noise and better constrained by the models \citep{Yue13,Yue16,Yue17}. Moreover, even after stacking of a few fields in C17, no significant cross correlation between CIB and CXB other than the [0.5-2] keV band was found. Is this due to the poor statistics in the harder X-ray bands or the astrophysical nature of the cross power emitter(s)? 

In this paper, we present the first measurements in the larger COSMOS field of the source-subtracted CIB and CXB fluctuations as well as their cross power by using the latest data from the $\spitzer$ Large Area Survey with the HyperSuprime-Cam \citep[SPLASH;][program ID 90042]{Takada10,Steinhardt14} and $\chandra$ COSMOS Legacy survey \citep[][program ID 901037]{Elvis09,Civano16}, which probes IR and X-ray power spectra to larger scales of $\sim$~3000$''$ and $\sim$~5000$''$, respectively. 

This paper is organized as follows. We describe the data reduction and analysis of the angular power spectra of our IR and X-ray observations in Section~\ref{s:irdata_map} \&~\ref{s:xdata_map}, respectively. In Section~\ref{s:cross} we present and analyze the cross power between the CIB and CXB fluctuations, followed by a discussion of the explanations of the cross power in Section~\ref{s:discussion}. Finally, a brief summary is in Section~\ref{s:summary}. 

\begin{table*}
  \centering 
  \caption{Analyzed IR Data}
  \label{tab:table1}
  \begin{tabular*}{1\textwidth}{cccccc}
    \hline
    Epoch & Dates &  number of AOR & number of 100s frames in 3.6\mum\  & number of 100s frames in 4.5\mum\ & total exposure time (h) \\
    \hline
     
1&   2013 Feb-Mar &   180  &  14697 &14698 &  408 \\          
2&   2013 Jul-Aug  &   182  &  14644 & 14648 &  407 \\			  
3&   2014 Feb-Mar &  185  & 10807 &10808&  300   \\                           
4&   2014 Jul-Aug  &   76  & 5647 & 5647  &  157  \\
    \hline
  \end{tabular*}
   \raggedright Notes:  Astronomical Observation Request = AOR. The mean depth of coverage is $\sim$ 5.1 h/pixel including all epochs and 4.4 h/pixel without epoch 4.
\end{table*}



\section{Near-Infrared Maps of the COSMOS field}\label{s:irdata_map}

\subsection{Data}\label{s:irdata}
The Infrared Array Camera \citep[IRAC;][]{Fazio04} is a four-channel camera, mounted on the\spitzer Space Telescope. IRAC has four different detectors, with size of 256$\times$256 pixels, a field of view of 5.2$'$$\times$5.2$'$ and FWHM of 1.6$''$--2.0$''$. The detectors simultaneously produce images at wavelengths of 3.6\mum, 4.5\mum, 5.8\mum\ and 8.0\mum. IRAC continued to provide images at 3.6\mum\ and 4.5\mum\ in the ``warm'' mission after the liquid helium cyogen runs out, but the 5.8 and 8.0\mum\ detectors are too warm to be useful for scientific observations.

Our SPLASH data are taken with the ``warm'' IRAC. SPLASH has obtained stacked images of two ultra-deep fields, COSMOS \citep[The Cosmic Evolution Survey;][]{Scoville07} and SXDS \citep[The Subaru/XMM-Newton Deep Survey; ][]{Ueda08} at both 3.6\mum\ and 4.5\mum\, with a total integration time of $\sim$2475 h, $\sim$1272 h of which was spent on the COSMOS. 

Located at Equatorial/Ecliptic/Galactic Coordinates of ($\alpha$, $\delta$) = (+150$^\circ$.12, +2$^\circ$.21), ($\lambda_{Ecl}$, $\beta_{Ecl}$) = (151$^\circ$.42, -9$^\circ$.36), ($l_{Gal}$, b$_{Gal}$) = (236$^\circ$.82, 42$^\circ$.12), COSMOS has the unprecedented combination of great depth and large sky coverage ($\sim$2.2 deg$^2$) and has been extensively observed at multiple wavelengths \citep[e.g.,][]{Werner04,Sanders07, Civano16}. In this work we use the SPLASH observations of the COSMOS field, which were taken at 4 different epochs from 2013 February to 2014 August, with intervals of $\sim$ 6 months (Table~\ref{tab:table1}). Such observation cadence is important to trace the seasonal contribution from zodiacal light to our measured CIB fluctuations \citep{Arendt16}. In all epochs, there is a blank area in the center of the field, with a size of $\sim$ 3.5$'$$\times$53.5$'$ ($\sim$~0.05~deg$^2$ area). This blank area was observed in the previous CANDELS survey \citep{Ashby15} and therefore was not re-observed in the shallower SPLASH survey.

\subsection{Data Reduction and Self-Calibration}\label{s:selfcalibration}
   \begin{figure*}
\centering
        \includegraphics[trim=50 100 50 100 clip, width=0.9\textwidth]{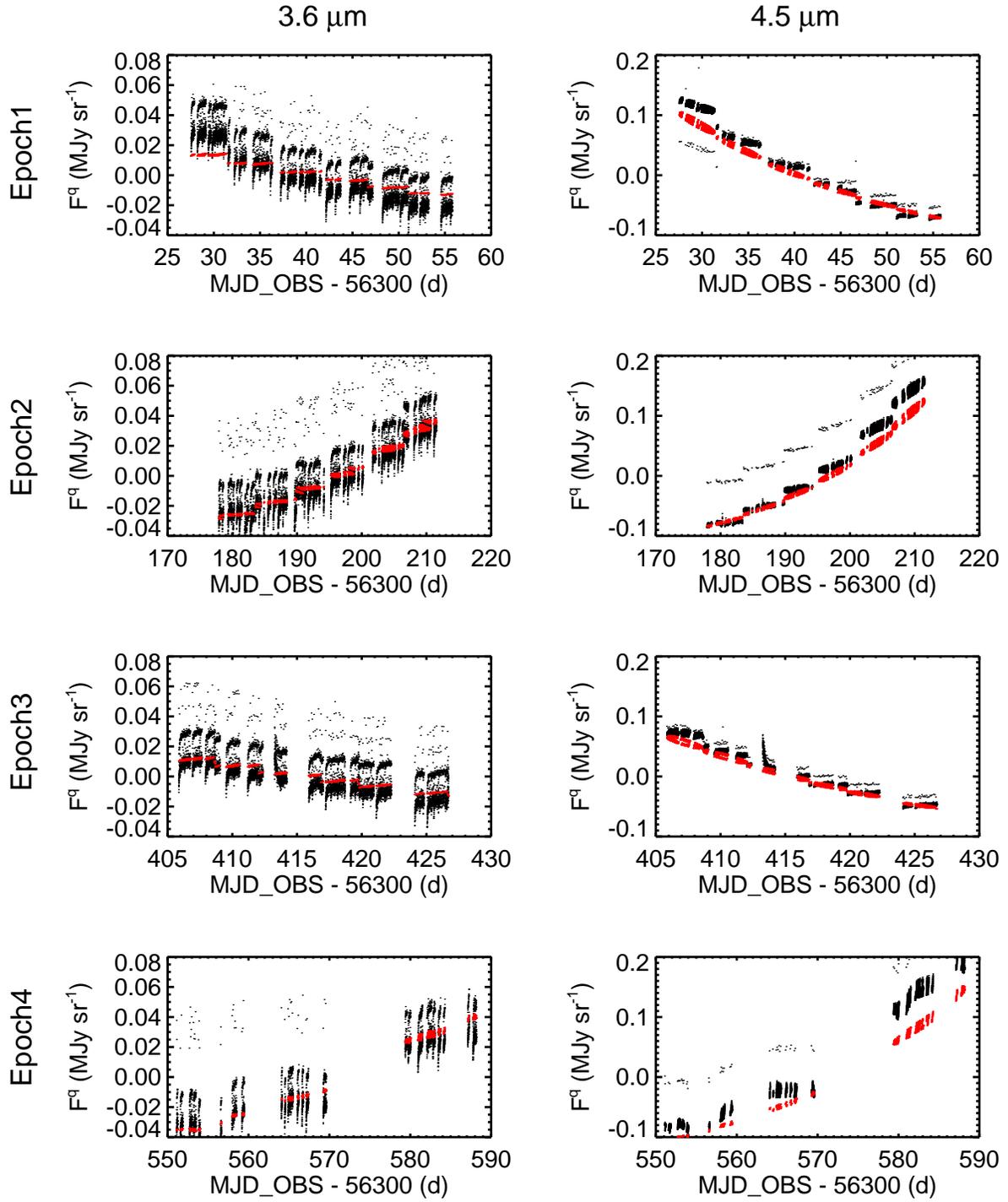}
    \caption{The $F^{q}$ offsets as a function of time, for each epoch and wavelength, derived by our self-calibration. All times in the figure are relative to a common starting date of MJD\_OBS[0]=56300. Black points show our derived F$^{q}$ values and red ones are the model predictions of the zodiacal light from the IRAC standard pipeline. The periodic outliers that appear above the bulk of the observations are caused by ``first frame effect'' (Section~\ref{s:selfcalibration}). The mismatch between F$^{q}$ and the model (especially shown in epoch 4 at 4.5\mum) suggests additional variations of the zodiacal light (Section~\ref{s:irmosaic}).  }{\label{fig:fq}}
      \end{figure*}

Our reduction starts with the corrected basic calibrated data (cBCD, see the $\spitzer$ Observer's Manual and the IRAC Data Handbook\footnote{\href{http://irsa.ipac.caltech.edu/data/SPITZER/docs/irac/iracinstrumenthandbook}{http://irsa.ipac.caltech.edu/data/SPITZER/docs/irac/iracinstrumenthandbook}}). The data were produced via the IRAC standard pipeline, which mitigated several common artifacts in the IRAC images that may impact the output images, such as ``column pull-down'' effects. 

For each wavelength, we first divide all the observed frames into 4 groups based on the four observation epochs. We further divide each epoch into 2 subsets based on their frame numbers, with ``A'' for odd frame IDs and ``B'' for even ones.  The difference maps between A and B, i.e., A-B maps (Section~\ref{s:irab}), can provide a robust estimate of the random instrumental noise, given that A and B subsets are only different by a mean interval of $\sim$100s and thereby share similar systematic errors. 
 
Analyzing the CIB fluctuations requires accurate measurements of the background signal and removal of the foreground emissions, which are relatively strong at the near-infrared and become challenging in the fluctuation analysis. The self-calibration algorithm by \citet{Arendt10} \citep[also see][]{Fixsen00} is a least-squares calibration method that can yield an optimal solution for both the calibration and the sky intensity.

We thereby process the 3.6\mum\ and the 4.5\mum\ data following the self-calibration method with modifications customized for our dataset, summarized as below.

Our self-calibration reconstructs the observed data by modeling the detector performance characteristics and the true sky intensity, using
 \begin{equation}\label{eq:model}
\rm \it D^i = S^{\alpha}+F^{p}+F^{q}
\end{equation}
where $D^i$ is the measured intensity of a single pixel $i$ from the single frame $q$, and $S^\alpha$ is the true sky intensity at location $\alpha$. $S^\alpha$ is what we need to derive here (i.e. the self-calibrated sky map). $F^p$ and $F^q$ describe the offset between the observed and the expected sky intensity, where $F^p$ remains constant with time and records the offset for detector pixel $p$, while $F^q$ is variable during the observations. 
For our IRAC dataset with fixed frame time, the detector dark current is incorporated in the $F^p$ term. Other time-dependent behavior of the detector can be described by $F^q$, as long as they can be characterized as a single value per readout per frame.  

One important component of $F^q$ that our calibration process takes into account is the zodiacal light (the scattered sunlight by the interplanetary dust, one of the strongest foreground light), which may affect frames from different epochs and lead to artificial fluctuations across the sky area. An important assumption of the self-calibration for the zodiacal light is that it is not spatially variant across the array and thus it can be described by $F^q$.

The model estimation for the zodiacal light is accessible from the header of the image of each frame via the keyword ZODY\_EST, which is estimated by the IRAC standard pipeline using the $\spitzer$ foreground model\footnote{\href{http://ssc.spitzer.caltech.edu/warmmission/propkit/som/bg/background.pdf}{http://ssc.spitzer.caltech.edu/warmmission/propkit/som/bg/background.pdf}}.

In Figure~\ref{fig:fq}, we present the $F^{q}$ offsets of our IRAC data which are variable during observations and compare them with the model predictions of zodiacal light. The 4.5\mum\ intensity varies with a larger amplitude of 0.3~MJy~sr$^{-1}$ than the 3.6\mum\ intensity varying within 0.1~MJy~sr$^{-1}$. There are periodic outliers which are far above the majority of the data. These outliers are caused by the ``first frame effect'', which is due to the fact that the detector response is noticeably different before and after long slews when the detector is periodically scanning the field. 
The longer the delay time between observations, the stronger this effect is. In addition, most of the $F^{q}$ show a good match with the modeled zodiacal light, indicating that the self-calibration process successfully recovers and thereby removes the zodiacal effect. However, the $F^{q}$ values at epoch 4 are significantly above the model, suggesting that the zodiacal light at this epoch has more complicated variations that cannot be fully described by the $F^{q}$ term, e.g., strong spatial gradient (Section~\ref{s:irmosaic})\citep{Arendt16}.

\begin{figure}
\centering
        \includegraphics[trim=180 140 180 150, clip, width=0.45\textwidth]{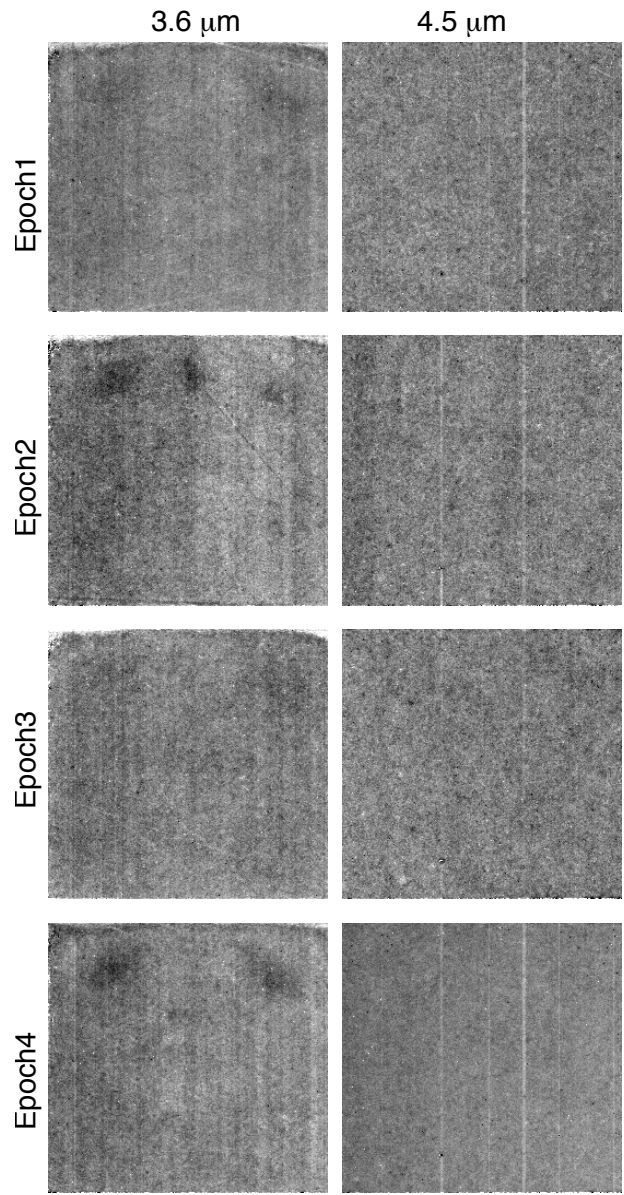}
    \caption{The $F^{p}$ maps (constant with time; see Equation~\ref{eq:model}), derived by self-calibration for each epoch (top to bottom) and wavelength (left: 3.6 \mum; right: 4.5\mum). All images are on a linear stretch of [-0.015,+0.015] MJy sr$^{-1}$. Multiple effects are identified and thereby removed by self-calibration, including, e.g., the dark patches at the top of some images produced by the residual stray light, and bright (white) vertical lines from ``column pull-down''.}{\label{fig:fp}}
      \end{figure}

In Figure~\ref{fig:fp}, we present the images of the detector offsets F$^{p}$ for 3.6 \mum~and 4.5\mum\ at different epochs. There are diffuse dark patches shown at the top of nearly all 3.6\mum\ images. These patterns are caused by residual stray light, produced by point sources and the diffuse background (e.g., zodiacal light) illuminating regions off the edges of the detector arrays. Although the standard IRAC pipeline attempted to correct for the stray light, we notice that the pipeline does not work as expected in the 3.6\mum\ especially at epoch 2 and epoch 4. Another notable feature is the vertical (bright) lines in the 4.5\mum\ images, appearing in the same column across multiple epochs. Those lines are ``column pull-down'' artifacts that being present in every frame taken, resulted from clusters of bad pixels at 4.5 \mum\ in the ``warm'' mission darks. There are some other visible features, e.g., the diagonal lines appearing in epoch1 of 3.6 \mum, which were generated by slews across bright sources. All of the artificial structures and patterns shown in the $F^{p}$ images are removed by our calibration process. 

The above procedure provides us with the expected sky intensities from all epochs for 3.6\mum\ and 4.5\mum. We then map these calibrated frames into a reference astrometry for making the mosaics.

   \begin{figure*}
   \centering
       \includegraphics[angle=90, trim=0 0 0 0, clip,width=1\textwidth]{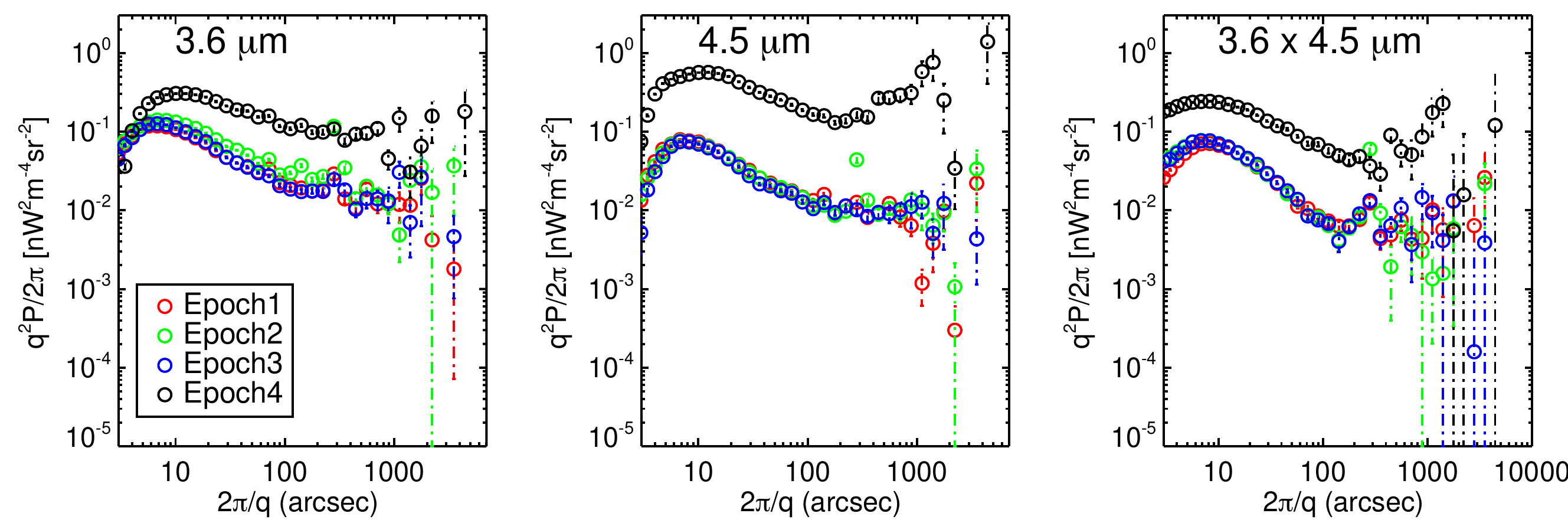}
    \caption{The auto and cross power spectrum for each epoch before stacking. Note that epoch 4 behaves very different from other epochs, which may be caused by additional variations of the zodiacal light at this epoch (Section~\ref{s:irmosaic}). The data from epoch 4 are not included in our further analysis (Section~\ref{s:irmosaic}).\label{fig:ir_eachepoch}}
      \end{figure*}
\begin{figure*}
\centering
       \includegraphics[trim=500 46 500 0, clip, width=0.45\textwidth]{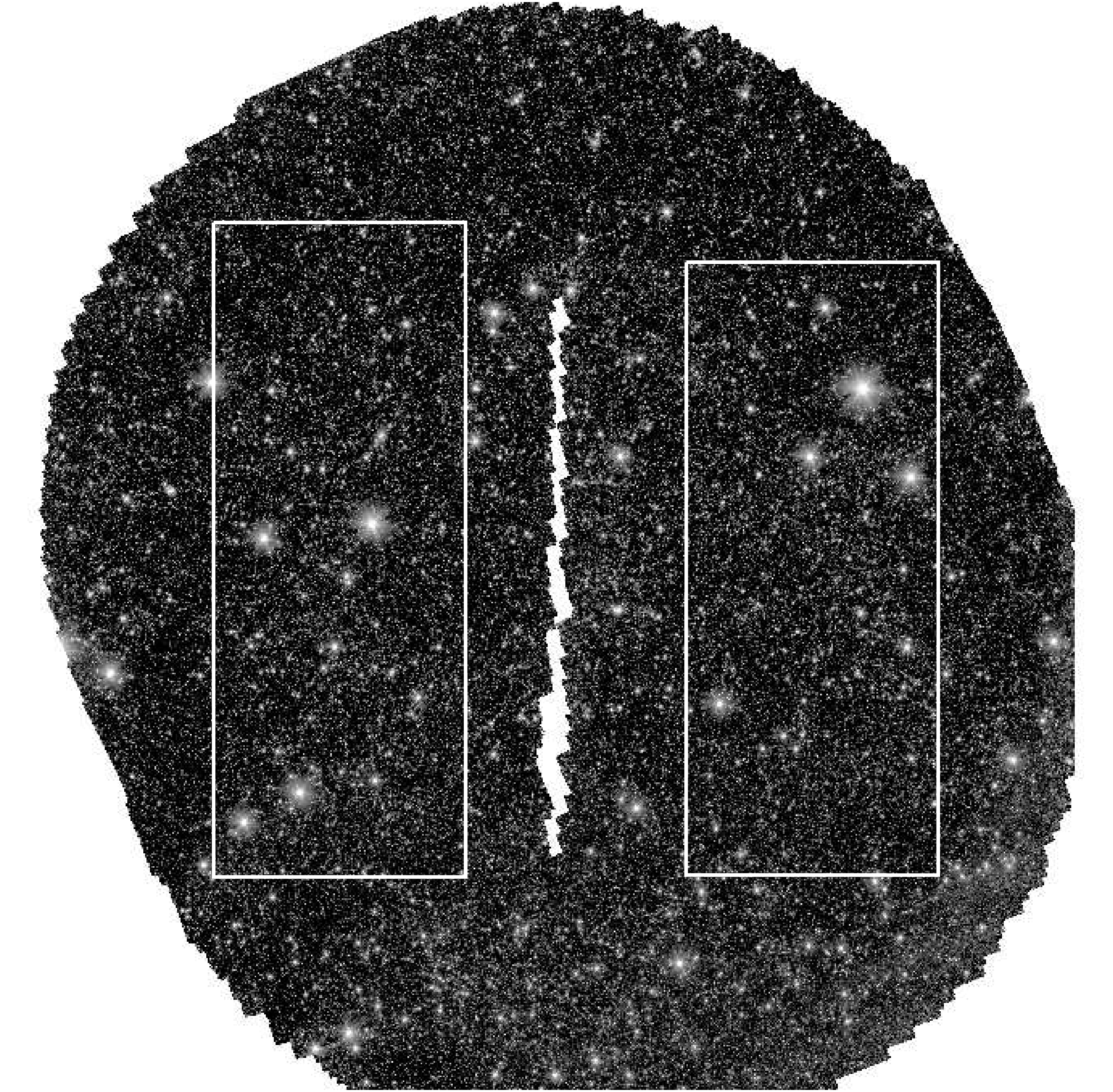}
         \includegraphics[trim=500 46 500 0, clip, width=0.45\textwidth]{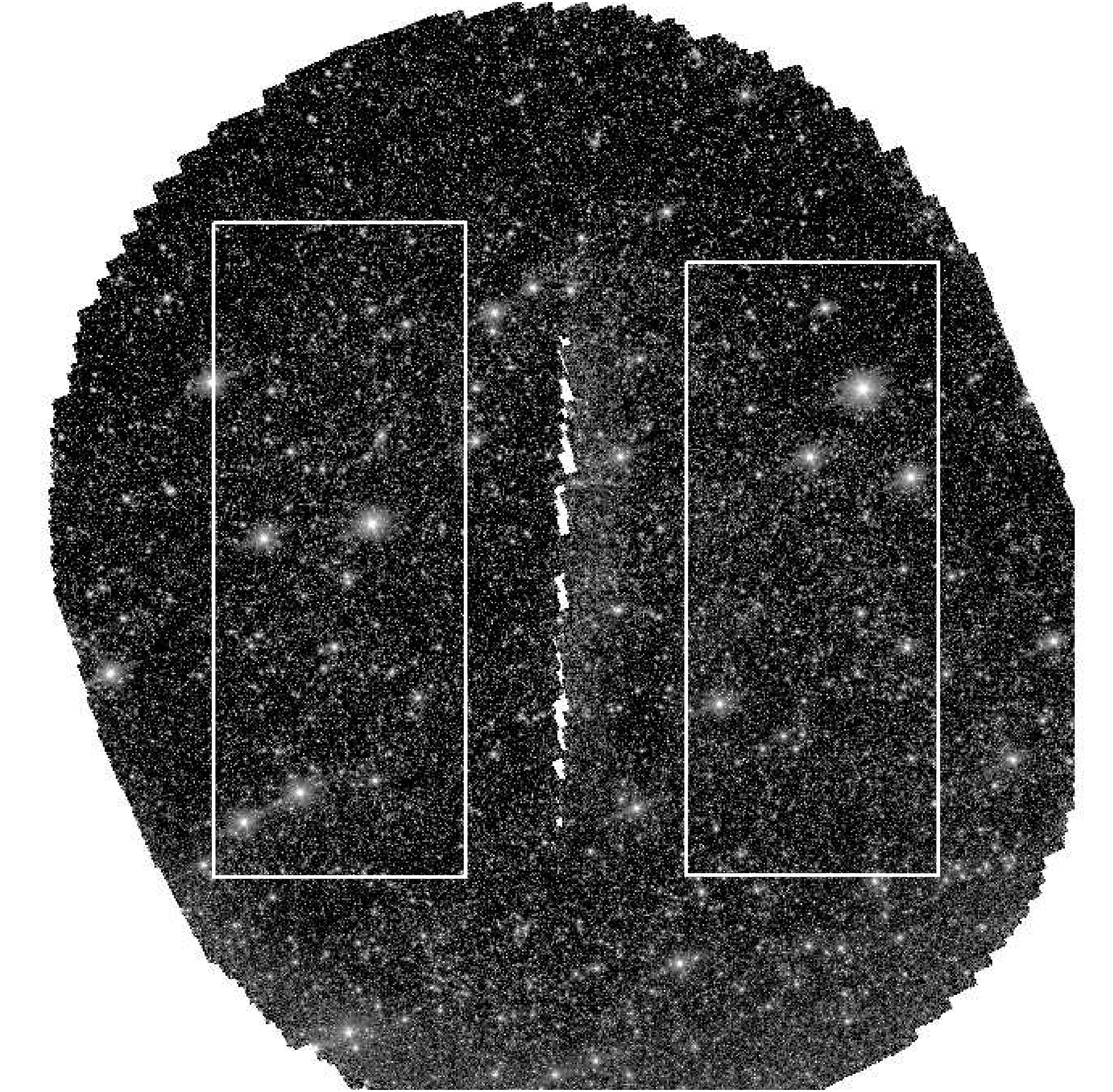}
 \caption{Mosaic sky maps, combining 3 epochs of data at 3.6\mum\ (left) and 4.5\mum\ (right) with orientation of North up and East left (Section~\ref{s:irmosaic}). The images are logarithmically scaled on the range~[1$\times$10$^{-4}$, 0.5] MJy sr$^{-1}$. The white rectangles show how we cut the whole field to two subfields, ``east'' ($\sim$~0.39 deg$\times$1.02 deg) and ``west'' ($\sim$~0.40 deg$\times$0.96 deg). \label{fig:irmap}}
\end{figure*}

\subsection{Mosaic Signal Maps}\label{s:irmosaic}
The self-calibrated frames are used to make mosaic maps at 3.6~\mum\ and 4.5\mum. 

There are some issues with the epoch 4 data. After subtracting a linear gradient from the mosaic maps of the 4 epochs, we find that the maps from the epoch 4 at both 3.6\mum\ and 4.5\mum\ show additional non-linear background gradients compared to other 3 epochs. By further examining the auto-power spectra (as shown in Figure~\ref{fig:ir_eachepoch}, techniques described in Section~\ref{s:irfft}), we find that the spectra from epoch 4 do not agree with others at both small and large scales. The discrepancy at small scales implies different shot-noise levels, resulting from the fact that epoch 4 data is shallower than other epochs by a factor of $\sim$2. 

The discrepancy at larger scales ($\gtrsim$200$''$) could be caused by a strong spatial variation of the zodiacal light during the epoch. We find that fitting a polynomial of degree 3 can remove the gradient, but it may also introduce some artificial structures when mosaicking maps from all epochs together. For the sake of the precision of our final results, we therefore only combine the images from the first 3 epochs into a mosaic map and exclude the epoch 4 data from our further analysis.

For the purpose of our subsequent power spectral analysis (Section~\ref{s:irfft}), we make the maps of both 3.6 and 4.5\mum\ to the same orientation (north up and east left). To avoid introducing artificial features and systematic errors into our maps and keep the consistency of our analysis, we leave the blank region in the center of the field  (Section~\ref{s:irdata}) and do not fill in with data from other surveys.
We select one rectangular subfield from the ``east'' and ``west'' part of the maps, respectively (Figure~\ref{fig:irmap}). Cutting the field into east/west subfield does not seriously affect the largest angular scale or the significance of our results, considering that the blank area is only  $\sim$3\% of the total area. 
 
After mosaicking all 3 epochs and cutting into two subfields, denoted as ``east'' and ``west'', we have two mosaic maps for each wavelength. We further rotate the standard beam map (PSF) for 3.6\mum\ and 4.5\mum\ to match each epoch of the mosaic maps, and then combined them to create the mosaic PSF for the map. These PSF maps are used for our subsequent modeling process (Section~\ref{s:irmask}). The resulting mosaic signal maps at 3.6 and 4.5\mum\ are shown in Figure~\ref{fig:irmap}. 

 \subsection{Mosaic Noise Maps}\label{s:irab}
In order to estimate the random noise of the maps, we divide the frames into the subset ``A'' with odd-numbered frame IDs and the subset ``B'' with even-numbered frame IDs. Following the same procedures described in Section~\ref{s:selfcalibration}, the COSMOS field from the ``A'' \& ``B'' subsets are cut to ``east'' and ``west'' rectangles with the same geometry as the signal map, i.e., A+B map (Section~\ref{s:irmosaic}). The noise map, which is the difference map between the self calibrated A and B images, i.e., A-B, should have no detected sources or instrumental systematics and only include random noise of the observation. We then use these noise maps to provide noise estimation in our subsequent analysis. 
   \begin{figure}
\centering
        \includegraphics[angle=90, trim= 0 0 0 0, clip,width=0.49\textwidth]{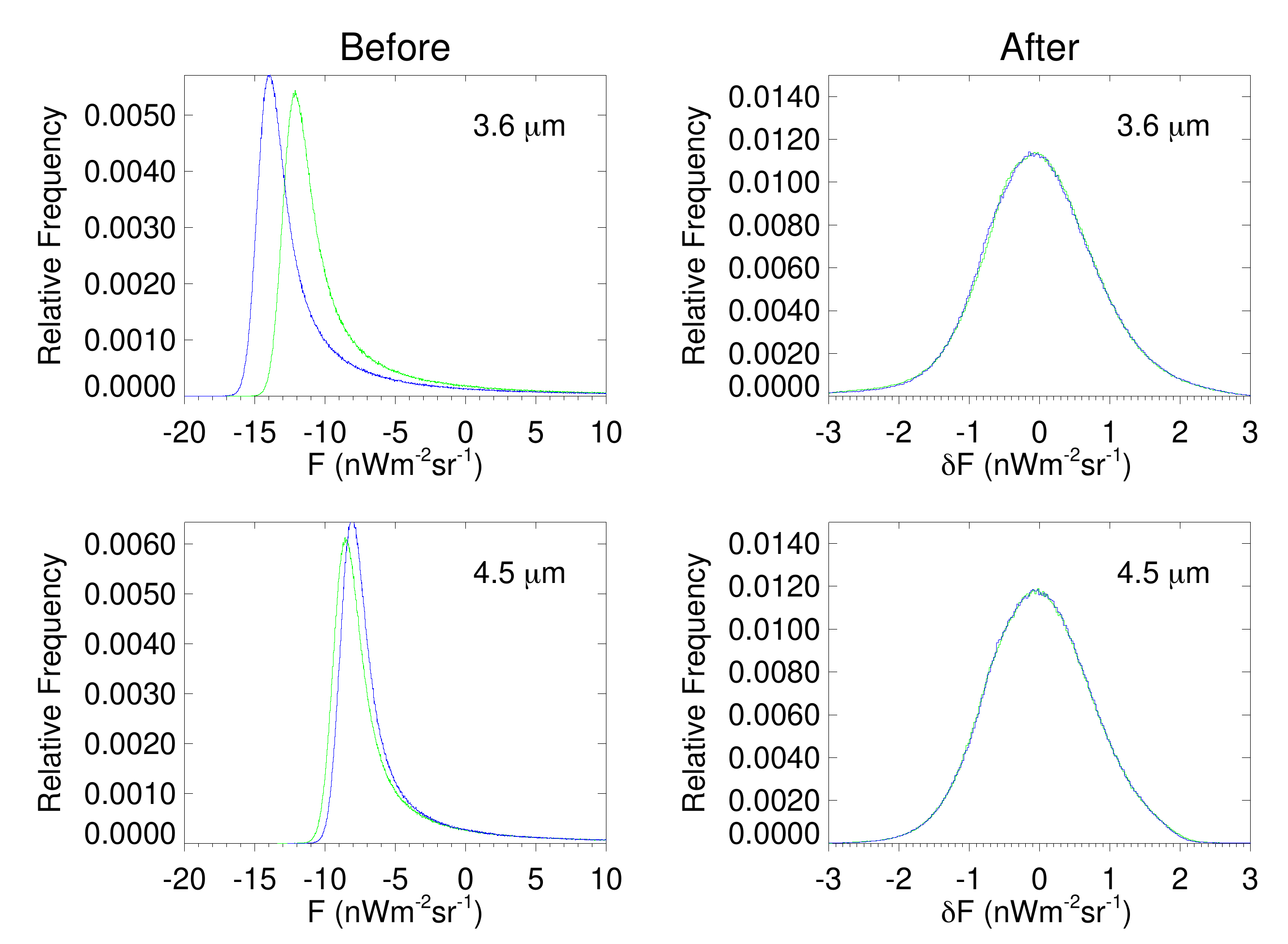}
    \caption{Histograms of the IR maps (subtracted with the mean values of unmasked pixels of the maps) before (left panels) and after masking (right panels), for 3.6 and 4.5\mum\ map, with green and blue color representing the east and west subfield.
    \label{fig:hist}}  
      \end{figure}
        \begin{figure}
\centering
        \includegraphics[angle=0, trim=10 0 0 5, clip,width=0.4\textwidth]{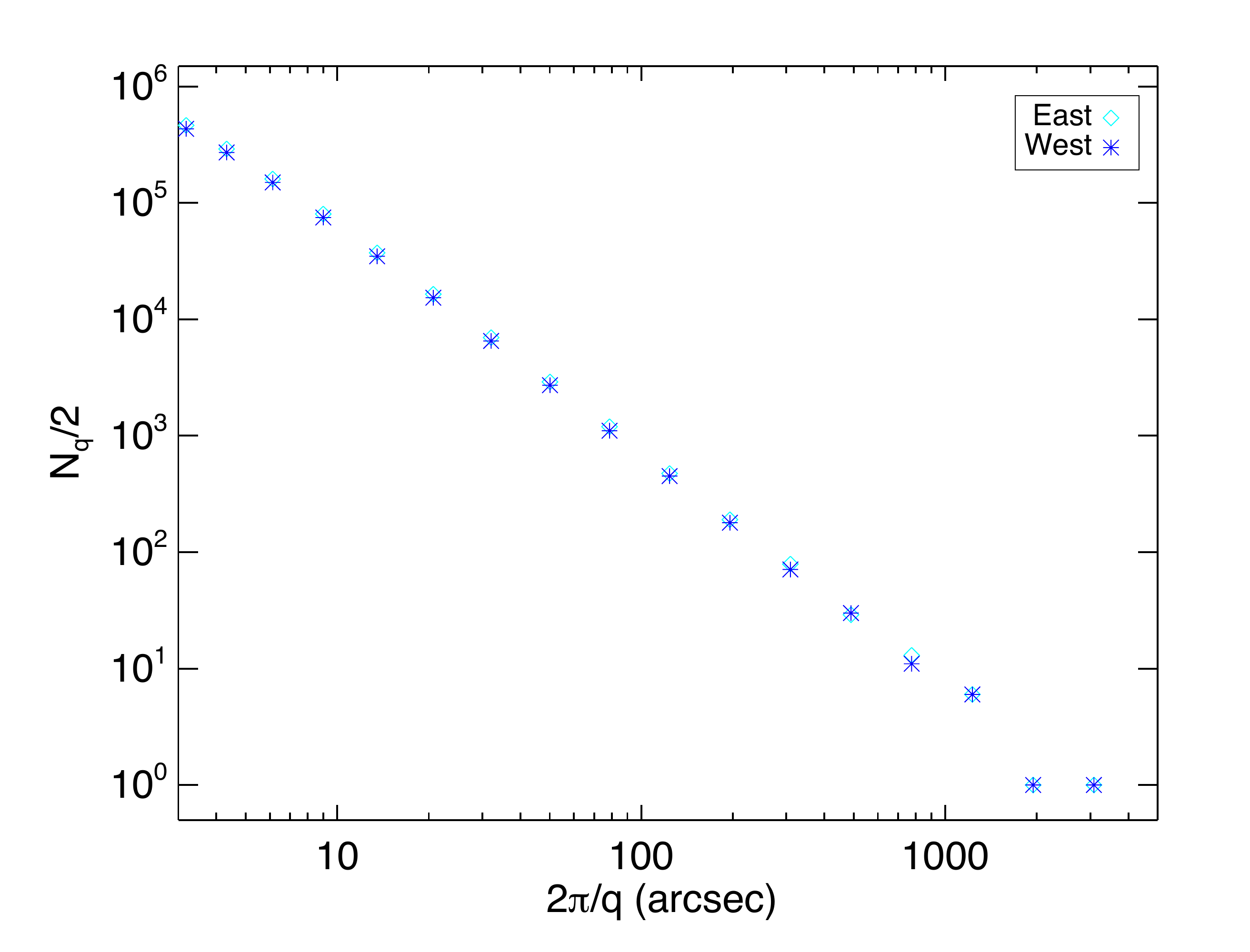}
    \caption{Number of independent Fourier elements that went into determining the power spectrum in our binning. Cyan and blue symbols are from the east and west subfield respectively. The same binning is applied to all of our Fourier analysis.  \label{fig:nq_f1}}  
      \end{figure}

\subsection{Source Removal and Masking}\label{s:irmask}
In order to investigate the CIB fluctuations from unresolved emissions, we need to first remove detected sources, such as galaxies and foreground stars, from the mosaic signal maps. This is mostly done by three steps in our analysis: (1) source modeling, to ensure the point sources and resolved extended sources are identified and removed from our signal maps, by following the recipe in \citet{Arendt10}; (2) masking, to clean up the modeling artifacts and the remaining emission from bright sources; (3) refining the source subtraction with extra masking.  

We use modeling with an iterative approach to identify and remove the detected sources. Our model assumes that the CIB and the underlying noise constitute a symmetric (if not Gaussian) intensity distribution, and resolved sources are distributed in the bright (positive) tail of the intensity distribution, causing a skewness. We thereby subtract the modeled sources from the maps, leading to trimming of the positive tail so that the overall intensity distribution is symmetric, without removing the CIB signals or digging into the noise. For each iteration, we synthesize a source at the brightest pixel in the image based on the mosaic PSF, created by combining the rotated PSFs from all epochs (Section~\ref{s:irmosaic}). We then subtract 60\% of this source intensity from the map to avoid over-subtraction, given that some extended sources could have very skewed and irregular profiles.

For each wavelength, we first carry out a sufficiently large number of iterations as to over-subtract the map. Then the results are reviewed to determine that actual number of iterations that minimizes skewness in the residuals and does not introduce any systematic effect. The final iteration number for 3.6 \mum\ east (west) and 4.5\mum\ east (west) is 272,000 (304,000) and 560,000 (384,000). 

The maps obtained from the above modeling process may still have residual intensity from the modeling artifacts at the locations of very bright sources, due to inaccurate point response functions and difficulty in modeling of sources with extended structures. Therefore in order to clean up those left-over emissions, we mask the image at an effective 3$\sigma$ surface brightness level and then expand each mask to include all the pixels neighbors within a square 3$\times$3 window. We also use SExtractor \citep{Bertin96} as an independent source detection technique. We set the source detection threshold at 3$\sigma$ level and then mask the sources detected by SExtractor. About 70\% of pixels are left with signal after subtracting the iterative models and masking from the signal maps. 

Finally, for the halo-like residuals around a couple of brightest stars on the maps, which may bring in some contaminations to our CIB fluctuation maps, we manually add a circular mask at the subtracted bright stars, with a radius of 170 pixels. We mask all the pixels within the circle that are brighter than the mean background level by $\gtrsim$3$\sigma$. In the end, our maps have $\approx$ 67\% pixels left for further analysis. 

In Figure~\ref{fig:hist}, we show the flux distributions of the mosaic maps in the east and west subfields before and after masking (with the fluctuations among the unmasked pixels). The two subfields share similar fluctuation distributions as expected. The final distributions are very close to a Gaussian profile after masking, with a mild skewness of $\lesssim$ 0.05.

   \begin{figure}
   \centering
       \includegraphics[trim=0 0 0 0, clip,width=0.51\textwidth]{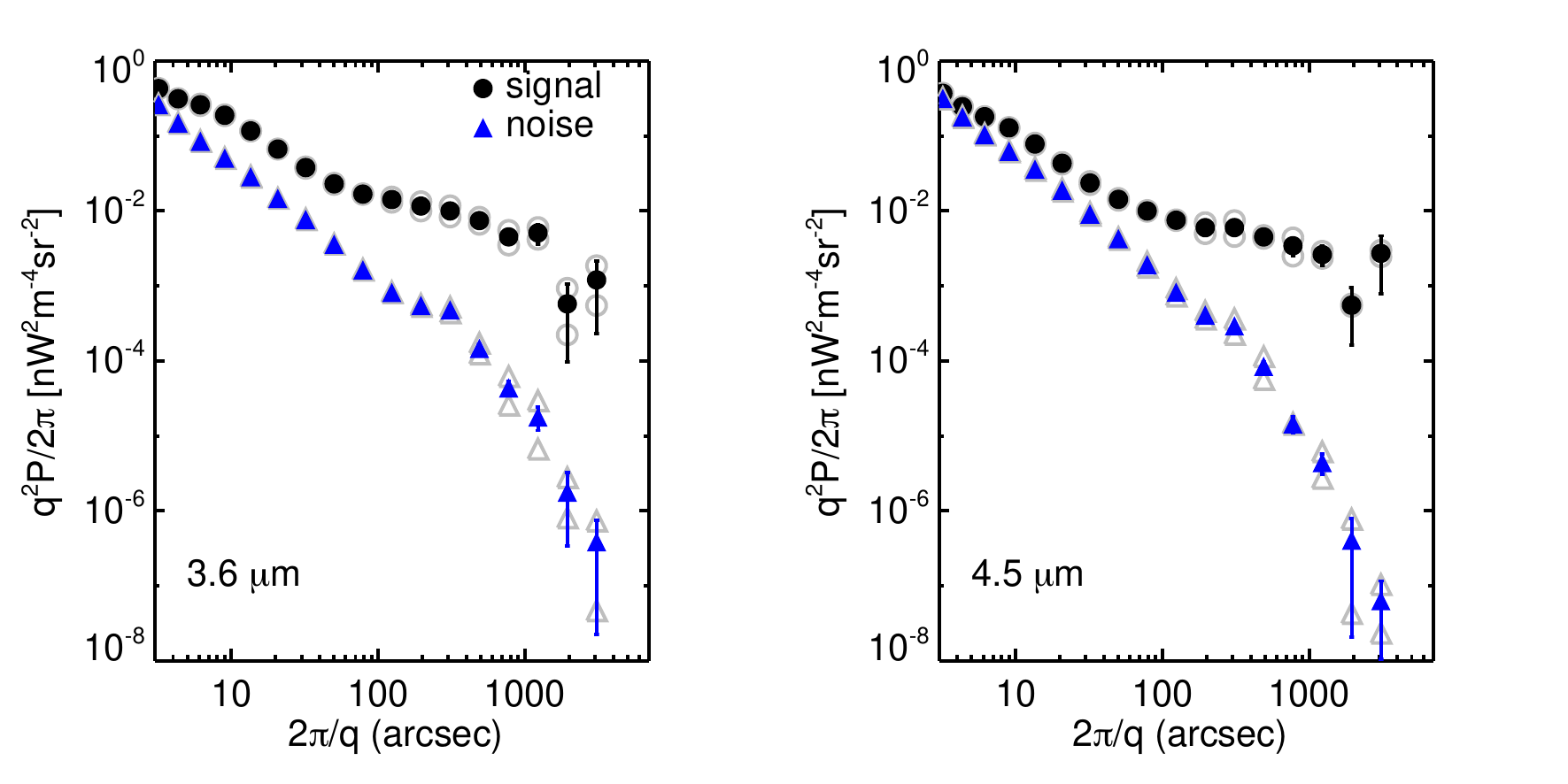}
    \caption{Top Panels: Fluctuations at 3.6\mum\ and 4.5\mum, from the signal map (black filled circles) and the noise maps (blue filled triangles). The gray open symbols are the fluctuations measured from the east and the west subfield. The measurements are consistent between these two subfields and the combined spectrum is the average of the two. \label{fig:ir_auto}}
      \end{figure}

  \begin{figure*}
   \centering
        \includegraphics[trim=0 0 0 0, clip,width=0.7\textwidth]{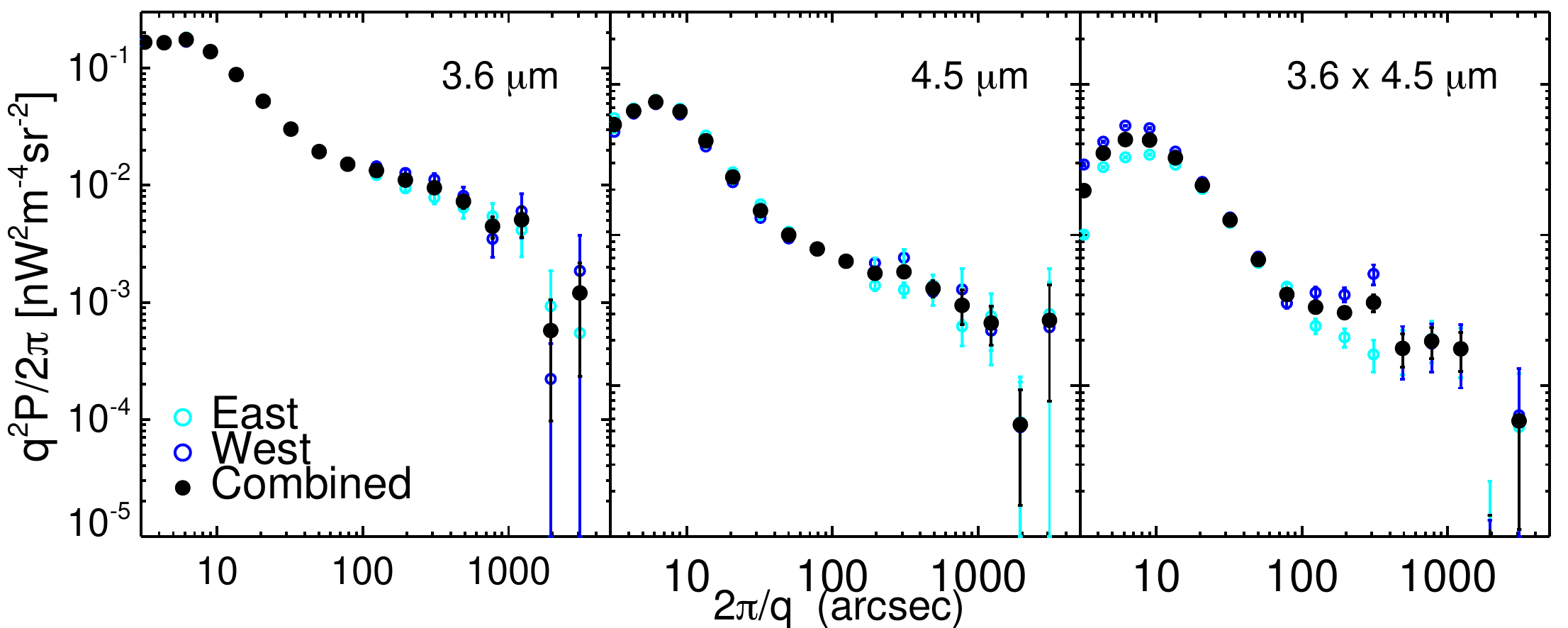}
    \caption{The clean auto power spectrum of 3.6\mum\ (left), 4.5\mum\ (middle) and the cross power spectrum (right). The measurements from subfields are shown with cyan and blue symbols and the combined values are shown as black filled circles.  }{\label{fig:ir_auto_cross}}
      \end{figure*}

\subsection{Fourier analysis of the fluctuation maps}\label{s:irfft}
We use the Fourier transforms (via a FFT function) of our observed maps to extract the power spectra of the fluctuations of the CIB.

We list in Table~\ref{tab:definitions} the definitions of several important quantities involved in our further Fourier analysis, following the conventions in \citet{Kashlinsky12} and \citet{Cappelluti13}. 

\begin{table*}
 \centering
\caption{Some important quantities involved in the Fourier analysis}\label{tab:definitions}
\begin{tabular}{p{3cm}p{5cm}p{8cm}}

 \hline
  Item                               &    Expression                         &   Description                         \\

 \hline 
  q	     &         					&Angular frequency    \\
   2$\pi$/q	     &         					& Angular scale   \\
N$_q$/2  & 									& The number of independent Fourier elements between q and q+dq \\
  $\delta$F(x)                   &         $I(x)$ - $\langle I(x)\rangle$                                   &   Fluctuation map                  \\
 $\Delta$(q) 		    & =$\frac{1}{4\pi^2}\int\delta F(x)exp(-ix \cdot q)d^{2}x$ & 2D Fourier Transform of the fluctuation map  \\
 P(q)  			&  =$\langle$$\mid$$\Delta(q)$$\mid$$^2$$\rangle$,~averaged over [q,q+$\delta$q] & Auto-power spectrum\\
$\sigma_{P(q)}$       & =$\frac{P(q)}{\sqrt{0.5N_q}}$ 			& The error estimation of P(q) \\
 $q^2~P(q)/2\pi$ &                & Mean squared fluctuations \\
P$_{1\times2}$(q) & = $\langle\Delta_{1}(q)\Delta^{*}_{2}(q)\rangle$ & Cross-power spectrum \\
$\sigma_{P_{1\times2}(q)}$ &  =$\sqrt{P_1(q)P_2(q)/N_q}$ & The error estimation of $P_{1\times2}(q)$  \\
${\it C}(q)$      & = $\frac{\mid P_{1\times2}(q)\mid^2}{P_1(q)P_2(q)}$ $\sim$  $\zeta_1^2$ $\zeta_2^2$  & Coherence, where $\zeta_1$ and $\zeta_2$ are the fractions of the emissions produced by the common population in the probed wavelengths \\ 
\hline
\end{tabular}
\end{table*}

In Figure~\ref{fig:nq_f1}, we show the number of independent Fourier elements that are included in determining the angular power spectrum from the east and the west subfields. In our study, we have $\sim$ 10$^6$ elements for both east and west subfields at the smallest scales, which is $\sim$1000 times more than previous studies with similar binnings \citep[e.g., Figure~1 in][]{Kashlinsky12} and enables an improved statistical significance in our IR auto power analysis. 

In Figure~\ref{fig:ir_auto}, we present the power spectra derived from the signal maps (i.e., A+B maps), the noise power spectra derived from the noise maps (i.e., A-B maps), and clean (noise subtracted) ones at 3.6\mum\ and 4.5\mum.  
We combine the power from both subfields based on their averages (shown as black filled circles). In Figure~\ref{fig:ir_auto_cross}, we present the clean auto power spectrum for each wavelength and the cross power spectrum between the two wavelengths.



  \section{X-ray Maps of the COSMOS Field}\label{s:xdata_map}
 
\subsection {$\chandra$ Data and reduction}
The $\chandra$ COSMOS Legacy survey covered the entire COSMOS field with an effective integration of 160 ks over the central 1.5 deg$^2$ area and 80 ks for the outer part of the field and a total observing time of 4.6 Ms. All observations were done in the VFAINT mode, which assures a low instrumental background level for our study. All of the 117 pointings of the\chandra observations over the COSMOS field are included in our study. Following \citet{Cappelluti13}, we reduce our data using the \chandra Interactive Analysis of Observations \citep[CIAO,][]{Fruscione06}, with the main procedures summarized as below.  

We cross match the $\chandra$ sources in the COSMOS fields with the optical catalogs of \citet{Capak07} and \citet{Ilbert09} to further improve the absolute pointing accuracy of our data. As suggested by the $\chandra$ X-ray Center documents, we then re-calibrated the event file with the level~1~products. Only the X-ray events with energy ranging from 0.3 keV to 10.0 keV recorded by the ACIS chips (with ccd\_id= 0, 1, 2, 3) are included in our analysis. We mask the bad events caused by, e.g, defective pixels, cosmic rays, or streak detections. We manually examine the light curve of the background and clean the background flares with \textsc{deflare}. The event files are further filtered with the GTI files. These background effects, if not removed, could contaminate the real cosmic background signal. The exposure maps 
are weighted sum for a range of energies rather than evaluated at a single energy value. Vignetting effects of the instruments are also taken into account. All of the above procedures are aimed to provide reliable measurements of the background emission for our further fluctuation analysis.
  
Additionally, in order to estimate the particle background (Section~\ref{s:xmosaic}), we apply similar procedures on the stowed data, which were taken when ACIS was ``stowed'' (e.g., the ACIS detector was out of the focal plane). They reflect the particle-only background. We merged all of the available stowed files from the \textsc{CALDB} library (version 4.7). The same \textsc{GAINFILE} is applied to both the stowed images and our observations. We also manually add the pointing (PNT) header keyword values in the background files which were originally set to zero. The \textsc{reproject\_events} tool is then used to handle the case of reprojecting the ACIS background file. Note that the same aspect solution file for each pointing is used here to guarantee that the stowed background is reprojected to the same focal plane as the source image.

\begin{figure*}
\centering
        \includegraphics[trim=550 50 550 0, clip,width=0.45\textwidth]{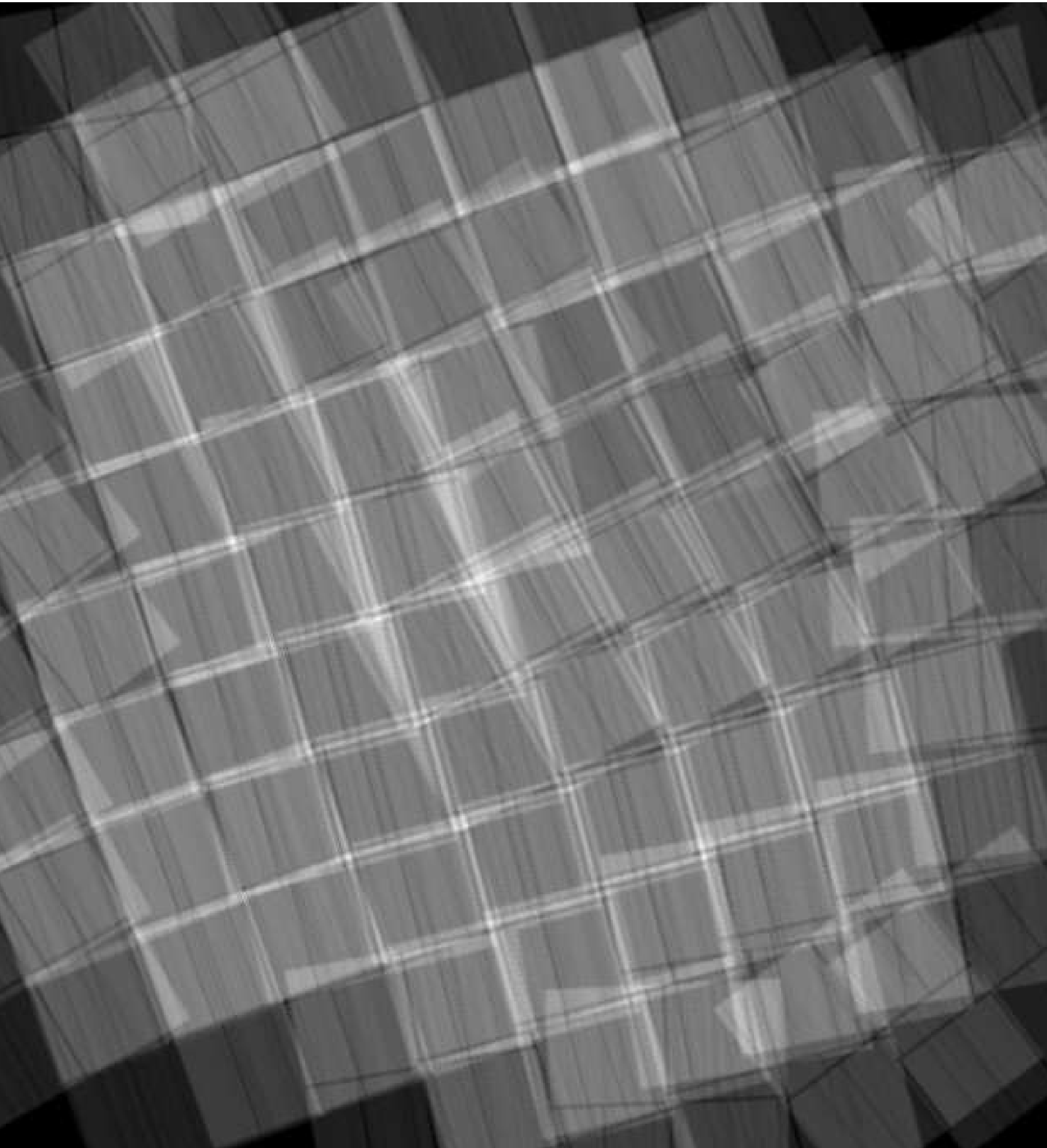}
         \includegraphics[trim=550 50 550 0, clip, width=0.45\textwidth]{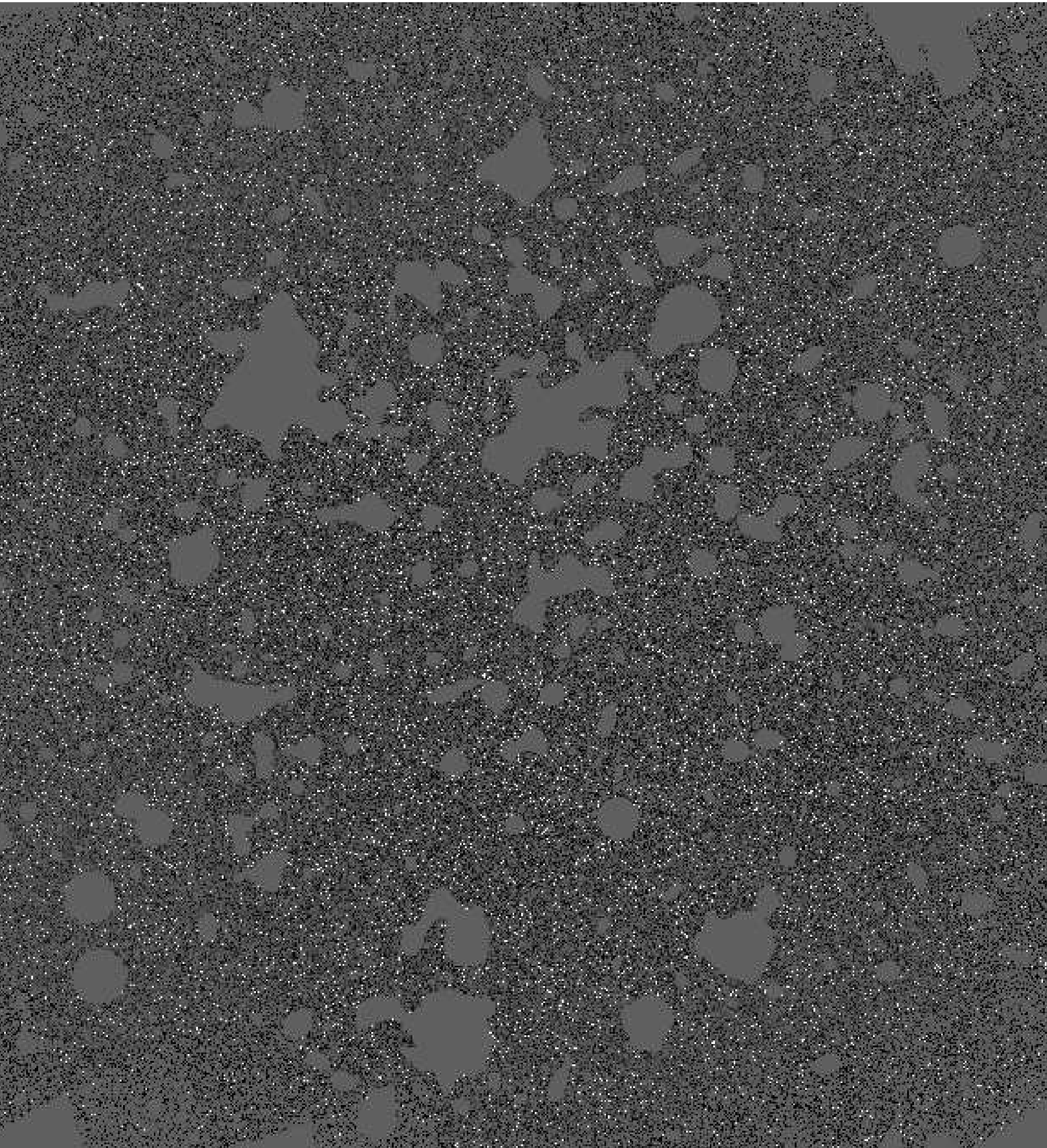}
    \caption{{\it Left:} Mosaic exposure map with all of the 117 pointings. {\it Right:} Mosaic fluctuation map at the soft X-ray band as an example. The size of the maps is $\sim$1.37 deg$\times$1.37 deg and the pixel scale is 0.984$''$/pixel. \label{fig:xmap} }
\end{figure*}

   \begin{figure*}
\centering
        \includegraphics[trim=0 0 0 0, clip, width=0.8\textwidth]{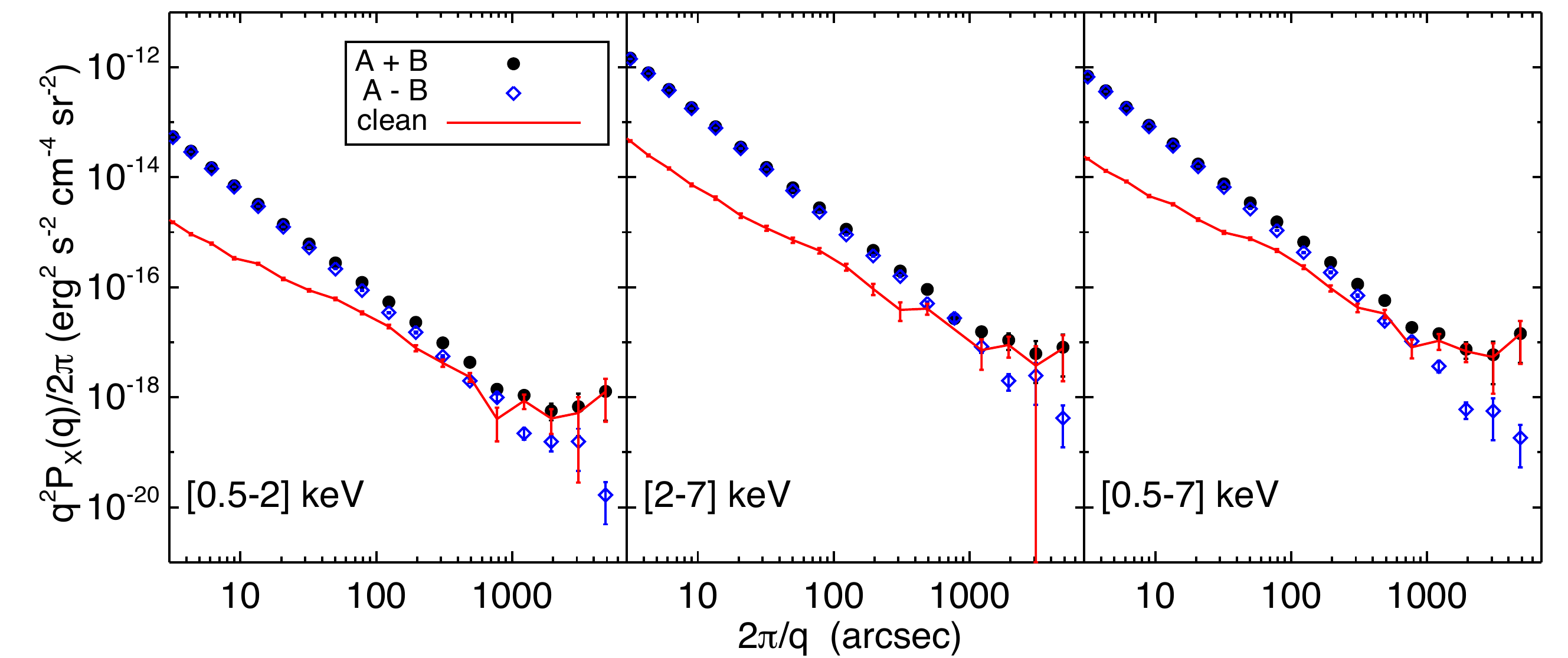}
    \caption{Auto power spectra for the soft, hard and broad band CXB fluctuation maps. The spectra from the signal maps are shown as black filled circles, ones from the noise maps are shown as blue diamonds and the difference of the two (the clean spectra) are shown in red.  \label{fig:autox}}  
      \end{figure*}

   \begin{figure*}
\centering
        \includegraphics[trim=0 0 0 0, clip, width=0.8\textwidth]{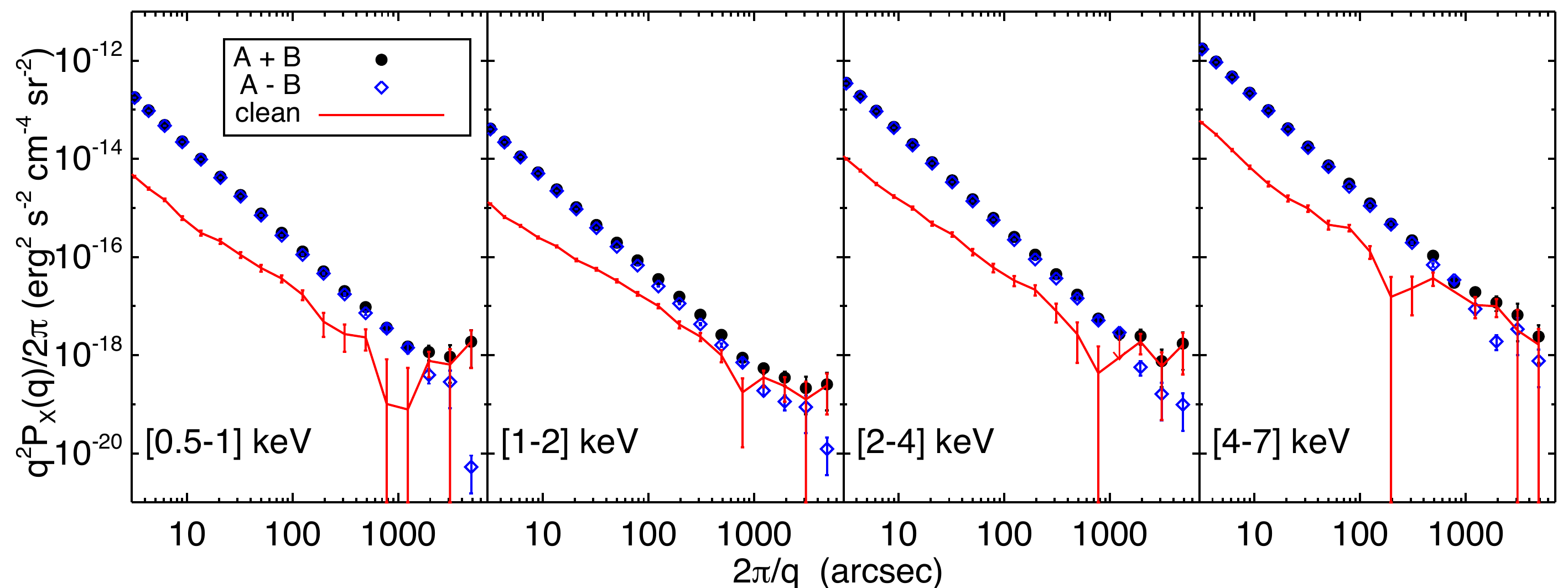}
    \caption{Auto power spectrum for each narrow X-ray band. The spectra from the signal maps are shown as black filled circles, ones from the noise maps are shown as blue diamonds and the difference of the two (the clean spectra) are shown in red.  \label{fig:autox_narrow}}  
      \end{figure*}

 \subsection{Mosaic maps and Fluctuation Maps}\label{s:xmosaic}
Similar to the IR map making (Section~\ref{s:irmosaic}), we create mosaic signal maps and noise maps for the X-ray analysis. As the X-ray photons are recorded not only by their energy but also their arrival time, they can be sorted by arriving time and grouped into image ``A'' with odd events and ``B'' with even events. A and B maps have the same exposure time and have been observed simultaneously so that the difference map of the two only contains instrumental effects. We then create the mosaic signal maps (A+B) and noise maps (A-B) to estimate the noise level. 

We create mosaic maps for different energy bands: [0.5-2]~keV(``soft''), [2-7]~keV (``hard'') and [0.5-7]~keV (``broad''), and further separate them to 4 narrow-width spectrum bands [0.5-1]~keV, [1-2]~keV, [2-4]~keV and [4-7]~keV. The first 3 bands are widely used in previous work and therefore necessary to compare our results, whereas the narrower bands are more useful for a coarse SED study of the cross power signal (Section~\ref{s:sed}).  

In order to create the X-ray fluctuation maps, we need to mask out resolved X-ray sources. We detect the X-ray sources based on the \citet{Civano16} catalog, including a total of 4016 X-ray sources detected down to flux limits of 2.2 $\times$ 10$^{-16}$, 1.5 $\times$ 10$^{-15}$, 8.9 $\times$ 10$^{-16}$ erg cm$^{-2}$ s$^{-1}$ in the soft, hard and broad band, respectively. A circular mask with a radius of 7$''$ is placed around each of the cross-matched sources between our maps \& the catalog. The mask radius is chosen to remove $>$90\% brightness of the detected X-ray sources. {\new We find that any escaped X-ray flux does not significantly contribute to the cross-power. }

The final X-ray masks are created by also incorporating the extended emission identified with groups and clusters of galaxies \citep{Finoguenov07}. As a result, the masking of X-ray sources leaves $\sim$ 85\% of the pixels for the CXB fluctuation analysis.

Another critical aspect in creating the background fluctuation maps is to appropriately remove the particle background component of the ACIS detectors from our signal maps. The particle background is due to high-energy particles hitting the detector CCDs and other surrounding materials and can be studied by stowing the ACIS inside the detector housing.\footnote{See notes in \href{http://cxc.cfa.harvard.edu/cal/Acis/Cal_prods/bkgrnd/current/}{http://cxc.cfa.harvard.edu/cal/Acis/Cal\_prods/bkgrnd/current/}}

\citet{Hickox06} first found that almost all of the high energy flux is from the quiescent particle background, due to the negligible\chandra effective area in the [9.5-12] keV range, and the particle background has a fairly flat spectral shape in the soft and hard band. We can get an accurate estimate of the particles by scaling the normalization of the stowed spectrum to the real observations by equating the [9.5-12] keV count rates. To be specific, the stowed images are rescaled to our data by the scaling factor which is the ratio between the total counts in the signal image and the stowed image with high energy in [9.5-12] keV. The scaled particle background is then subtracted from the signal maps. The counts of the resulting signal maps were then renormalized using the exposure maps to create an exposure-corrected cosmic background.

We then performed a Monte Carlo simulation on the resulted signal maps and stowed maps with 1000 Poissonian realizations of each of the two backgrounds. During each realization, a mean background is also calculated by distributing the total counts in the signal maps observation maps outside of the mask (i.e., unmasked pixels), following the patterns in the exposure map. Our final fluctuation maps are then created by subtracting the mean background from the signal background. For each of the 7 X-ray bands, we create a source subtracted fluctuation map. As an example, in Figure~\ref{fig:xmap}, we show the map of the soft band and the effective exposure map.

  \subsection{Power Spectra}\label{s:xps}
Using the fluctuation maps, we then extract the power spectra following similar methods as described in Section~\ref{s:irfft}. In Figure~\ref{fig:autox} and Figure~\ref{fig:autox_narrow}, we present the auto power spectra for signal and noise maps, as well as the difference of these 2 power spectra.

Moreover, in order to compare our results to previous studies, which mostly adopted the division of 20$''$, we focus on the auto- and cross-power analysis above 20$''$ as well. In Table~\ref{tab:xproperties}, we summarize the X-ray photon counts before and after applying the X-ray masks, as well as the amplitude for the auto power spectra.

\begin{table}
  \centering
\caption{X-Ray Map Properties}
  \label{tab:xproperties}
  \begin{tabular}{llllc}
\hline
Band    & N$_{cts}^b$  & N$_{cts}^a$ &     N/pixel$^a$  & $\langle P_X \rangle$  \\
 \hline
 
    {[0.5-1]} keV&     241464 &    164981 &      0.01 &                 0.42 $\pm$ 0.03   \\
    {[1-2]} keV &     541166 &    335080 &      0.01 &                 0.19 $\pm$ 0.01   \\
    {[2-4]} keV &     760227 &    569329 &      0.02 &                 0.96 $\pm$ 0.06   \\
    {[4-7]} keV &     858569 &    680262 &      0.03 &                 3.37$ \pm$ 0.28  \\
    
    {[0.5-2]} keV &     782630 &    512965 &      0.02 &                 0.32 $\pm$ 0.01   \\  
    {[2-7]} keV &    1618796 &   1279179 &      0.05 &                 4.44$\pm$ 0.23   \\ 
    {[0.5-7]}keV &    2401426 &    1792144 &      0.07 &                 3.88$\pm$ 0.11   \\  

\hline
\end{tabular}
   \raggedright Notes: for each band, we summarize the X-ray photon counts before masking (N$_{\rm cts}$$^{\rm b}$) and after masking (N$_{\rm cts}$$^{\rm a}$), counts/pixel after masking (N/pixel$^{\rm a}$) and the auto-power spectrum amplitude ($P_{X}$) averaged at angular scales beyond 20$''$, in units of 10$^{-24}$ erg$^2$ cm$^{-4}$s$^{-2}$sr$^{-1}$. 
   \end{table}

   \begin{figure}
\centering
               \includegraphics[trim=5 10 15 10, clip, width=0.4\textwidth]{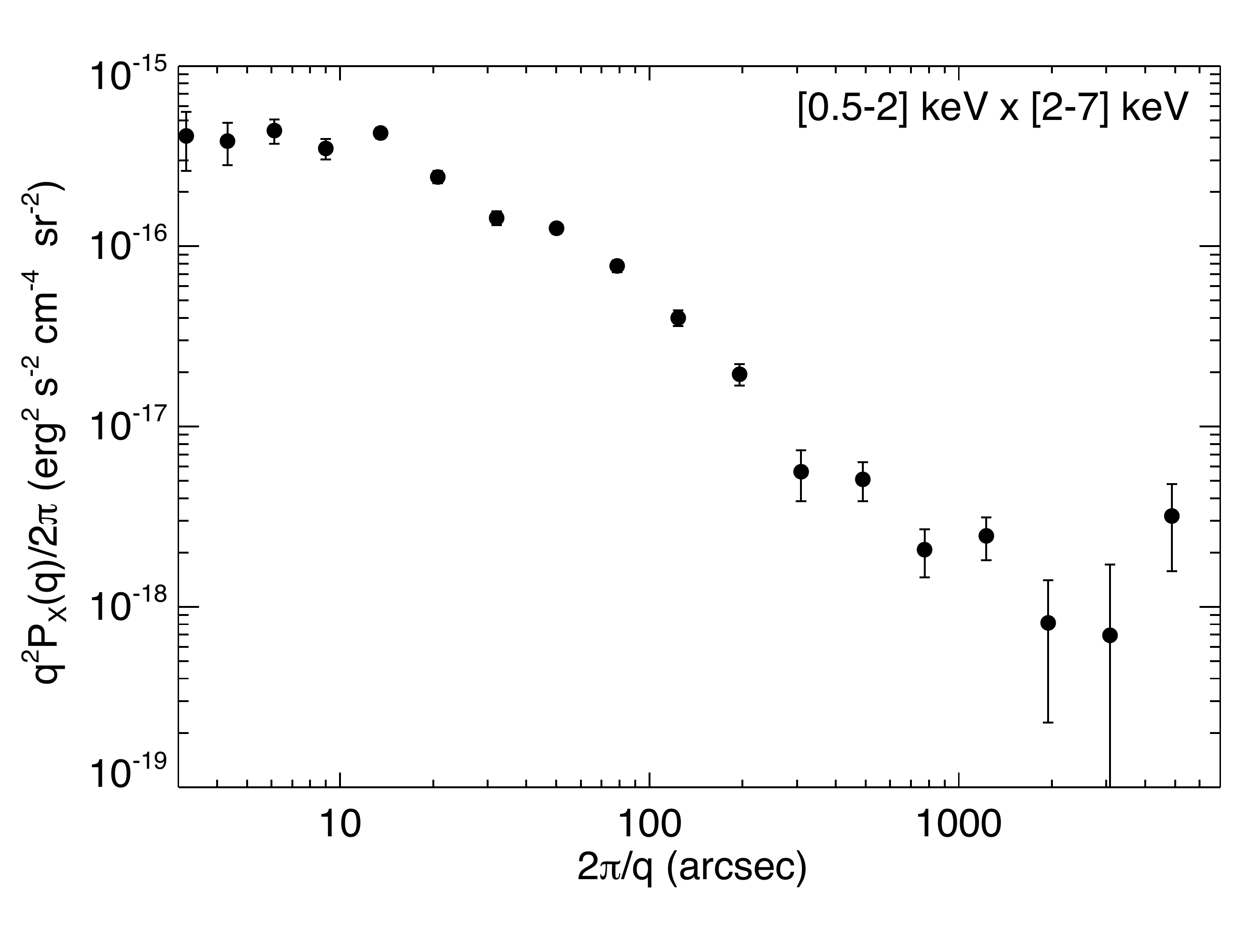}
    \caption{The cross power spectrum between [0.5-2] \& [2-7] keV, with the largest scale around~$\sim$ 5000$''$. The average coherence of the two bands is $\sim$ 0.1. \label{fig:crossx}}  
      \end{figure}

   \begin{figure*}
\centering
               \includegraphics[trim=0 0 0 0, clip, width=0.7\textwidth]{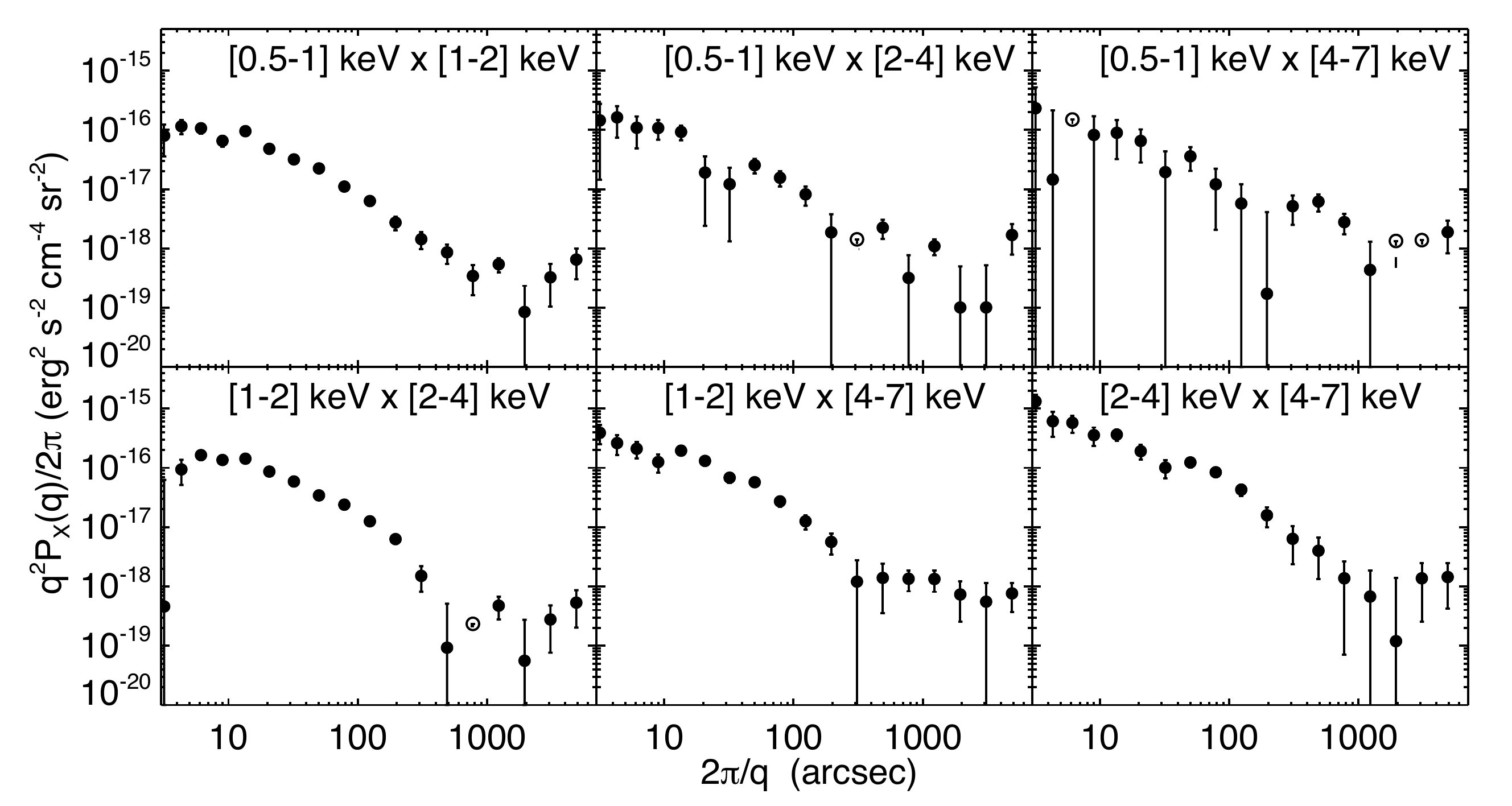}
    \caption{The cross power spectra between narrow X-ray bands, with the largest angular scale around $\sim$ 5000$''$. The absolute values of the negative power are denoted with open circles.  \label{fig:crossx_narrow}}  
      \end{figure*}

\begin{table*}
  \centering
  \caption{Cross-Power-Spectrum Amplitude $>$ 20$''$ in units of 10$^{-19}$ erg s$^{-1}$ cm$^{-2}$ nW m$^{-2}$ sr$^{-1}$}
  \label{tab:irx}
 \begin{tabular*}{1\textwidth}{l @{\extracolsep{\fill}} c c c c c c }
    \hline
 & \multicolumn{2}{c}{[0.5-2] keV} &  \multicolumn{2}{c}{[2-7] keV} &  \multicolumn{2}{c}{[0.5-7] keV}  \\
\cline{2-3}
\cline{4-5}
\cline{6-7}
     & $\langle P_{IR,X} \rangle$ & $\langle P_{IR,A-B}\rangle$ &  $ \langle P_{IR,X}\rangle $& $\langle P_{IR,A-B} \rangle$ & $ \langle P_{IR,X}\rangle$ & $\langle P_{IR,A-B} \rangle$ \\
         \hline
         
  3.6\mum~ &   1.20 $\pm$ 0.31 (3.9 $\sigma$)  & 0.46 $\pm$ 0.30  &    3.51 $\pm$ 1.61 (2.2 $\sigma$) & -0.44 $\pm$ 1.58 &  4.26 $\pm$ 1.11 (3.8$\sigma$) & 0.59 $\pm$ 1.08 \\
  
   4.5 \mum\ ~&   1.00 $\pm$ 0.25 (4.1 $\sigma$)  &0.07 $\pm$ 0.24  &   4.90 $\pm$ 1.28 (3.8 $\sigma$) & 1.27 $\pm$ 1.26 &  4.73$ \pm$ 0.88 (5.4 $\sigma$) & 0.85$ \pm$ 0.86 \\
  3.6+4.5 \mum ~&   1.08 $\pm$ 0.19 (5.6 $\sigma$)  & 0.22 $\pm$ 0.19 &    4.36 $\pm$ 1.00 (4.4 $\sigma$) & 0.61$ \pm$ 0.99 &  4.55 $\pm $0.69 (6.6 $\sigma$) & 0.75$\pm$ 0.67 \\

     \hline
    
  \end{tabular*}
  \end{table*}

In order to examine if the background emission in different X-ray energy bands are dominated by the same types of astrophysical sources, we thereby compute the cross power between each pair of the images (with independent energy bands). All of the power spectra reach an angular scale of $\sim$ 5000$''$ with high accuracy. The cross power spectrum between the soft and hard X-ray band is shown in Figure~\ref{fig:crossx}. The cross power reaches a lowest value around $~$2000$''$ and the values appear to be increasing afterwards.

As defined in Table~\ref{tab:definitions}, we calculate the coherence to learn about how much of the X-ray emission in the cross-power spectrum is produced by the common population. As a result, we find that the {\new mean coherence over all angular scales above 20$''$} for soft-hard band is 0.10, meaning that soft and hard power is barely correlated and only $\sim$~32\% of the X-ray sources are common to the both bands.  {\new Note that, however, the coherence here (also in the following analysis) is a lower limit as the X-ray maps are dominated by shot-noise, which leads to overall lower coherence.} The cross power spectra for narrow X-ray bands are shown in Figure~\ref{fig:crossx_narrow}.



 \section{CIB and CXB cross power}\label{s:cross}
 
 \subsection{COSMOS}\label{s:cosmos}
    \begin{figure*}
\centering
      \includegraphics[width=1\textwidth]{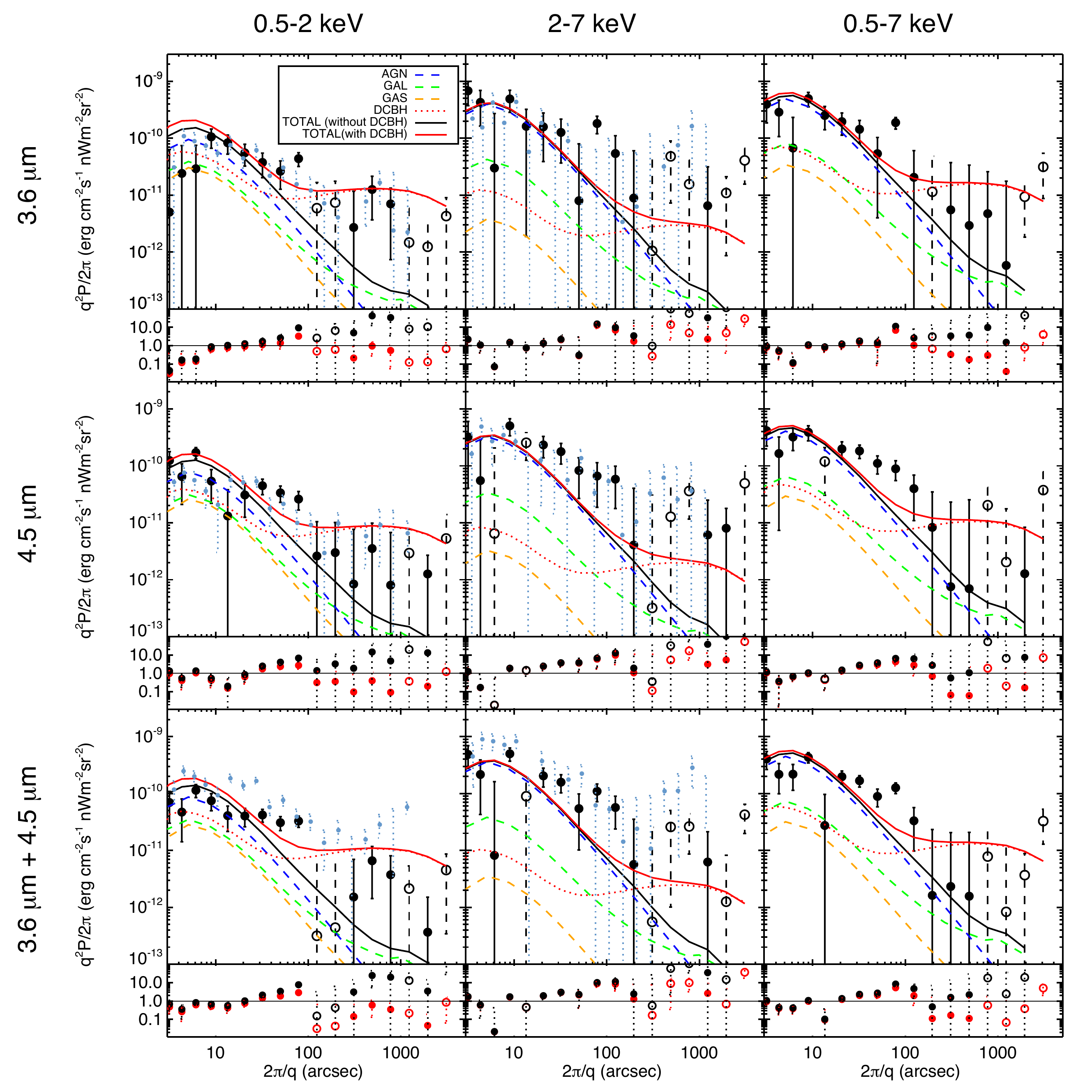}
    \caption{CIB-CXB cross-power spectra with COSMOS data: between 3.6\mum\ and 0.5-2 keV, 2-7 keV and 0.5-7 keV (Top panels); between 4.5\mum\ and CXB (Middle panels); combined CIB (3.6\mum\ + 4.5\mum) vs. CXB (Bottom panels). Open circles with dashed error bars denote the absolute values of negative results. Our measurements are mostly consistent with previous studies in C17 (shown as blue symbols, the stacked results are shown in Figure~\ref{fig:irx_stack}). The black solid line shows the sum of the contribution from known (unresolved) $z$$<$6 populations: AGN (blue dashed line), star-forming galaxies (green dashed line) and hot gas in clusters (orange dashed line). The predicted signal from DCBHs \citep[][private communication for revised lines]{Yue13} is shown with red dotted line. The red solid line shows the total signal from DCBHs (plus remaining known populations), which is the reason why it may have slightly higher shot-noise than some of the measurements. At the bottom of each panel, the ratio between the measurements (black circles) and the model without DCBHs (black solid lines) are plotted with black symbols, while the ratio between the measurements (black circles) and the model with DCBHs (red solid lines) are plotted with red symbols. The mean powers are shown in Table~\ref{tab:irx}.}{\label{fig:irx}}
      \end{figure*}
    \begin{figure*}
\centering
        \includegraphics[width=1\textwidth]{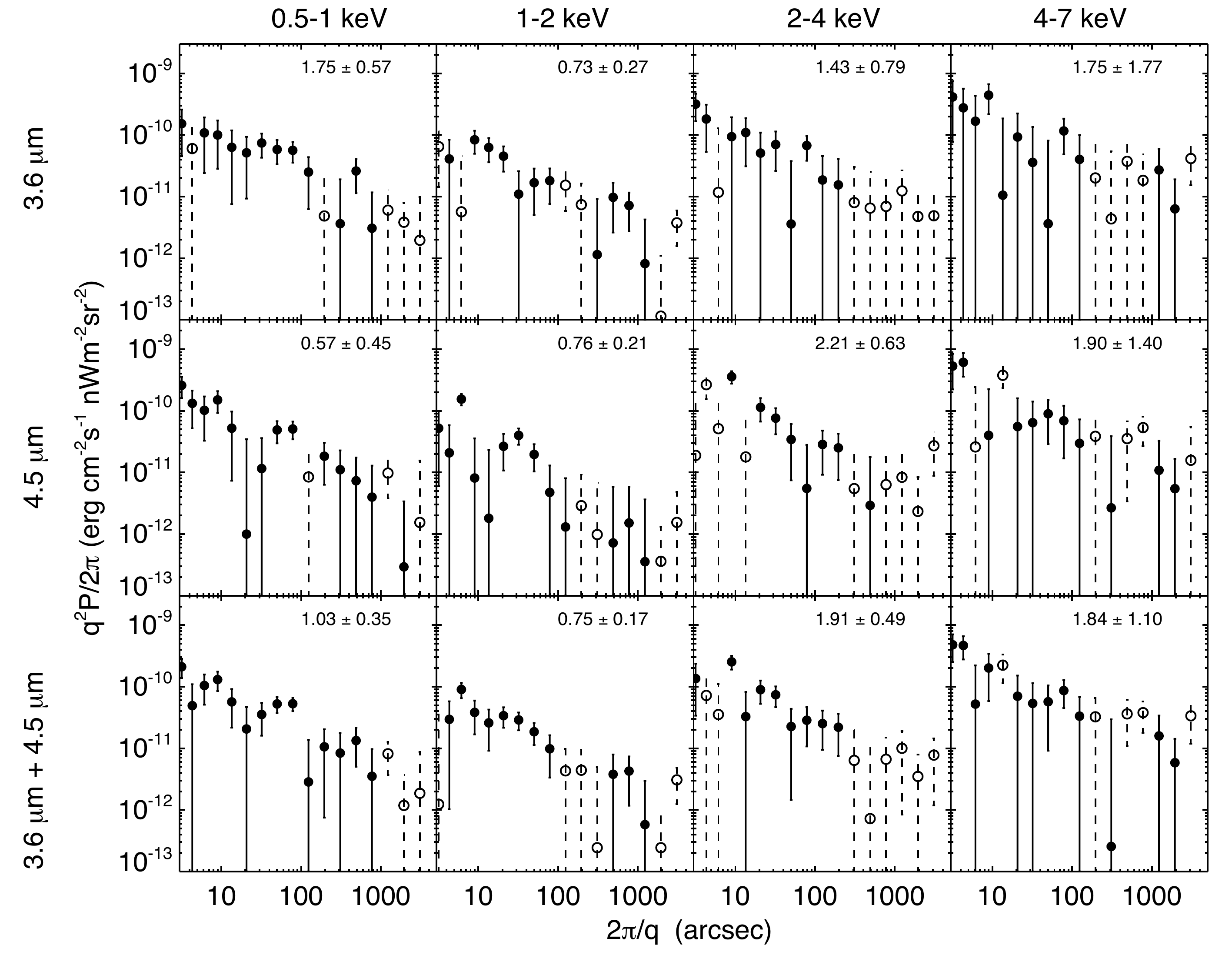}
    \caption{{\it Top panels:} CIB-CXB cross-power spectra: between 3.6\mum\ and [0.5-1] keV, [1-2] keV,  [2-4] keV and [4-7] keV,  {\it Middle panels:} with 4.5\mum, {\it Bottom panels: }CXB vs combined CIB (3.6\mum\ + 4.5\mum). Open circles with dashed error bars denote the absolute values of negative results. The mean power (calculated above 20$''$) and error are shown in the corner of each panel, in units of 10$^{-19}$~erg~s$^{-1}$~cm$^{-2}$~nW~m$^{-2}$~sr$^{-1}$.}{\label{fig:irx_narrow}}
      \end{figure*}

At this point, we have 4 infrared fluctuation maps (at 3.6 \& 4.5\mum) for the east and west subfield and 7 X-ray fluctuation maps for the entire COSMOS field. In order to investigate the contribution from possible X-ray emitters to the CIB excess fluctuations, we need to study the cross power between the CIB and CXB fluctuation maps. 

{\new For computational convenience, we match the X-ray maps to the IR maps, so that they have the same astrometry and pixel size (1.2$''$/pixel). }We merge the IR and X-ray mask and obtain the combined mask, M$_{IR, X}$, by multiplying the X-ray mask, M$_X$, and the IR masks, M$_{IR}$, and we have $\sim$60\% unmasked pixels after applying M$_{IR, X}$ for further analysis.

   \begin{figure*}
\centering
        \includegraphics[trim=30 20 0 20, clip, width=0.8\textwidth]{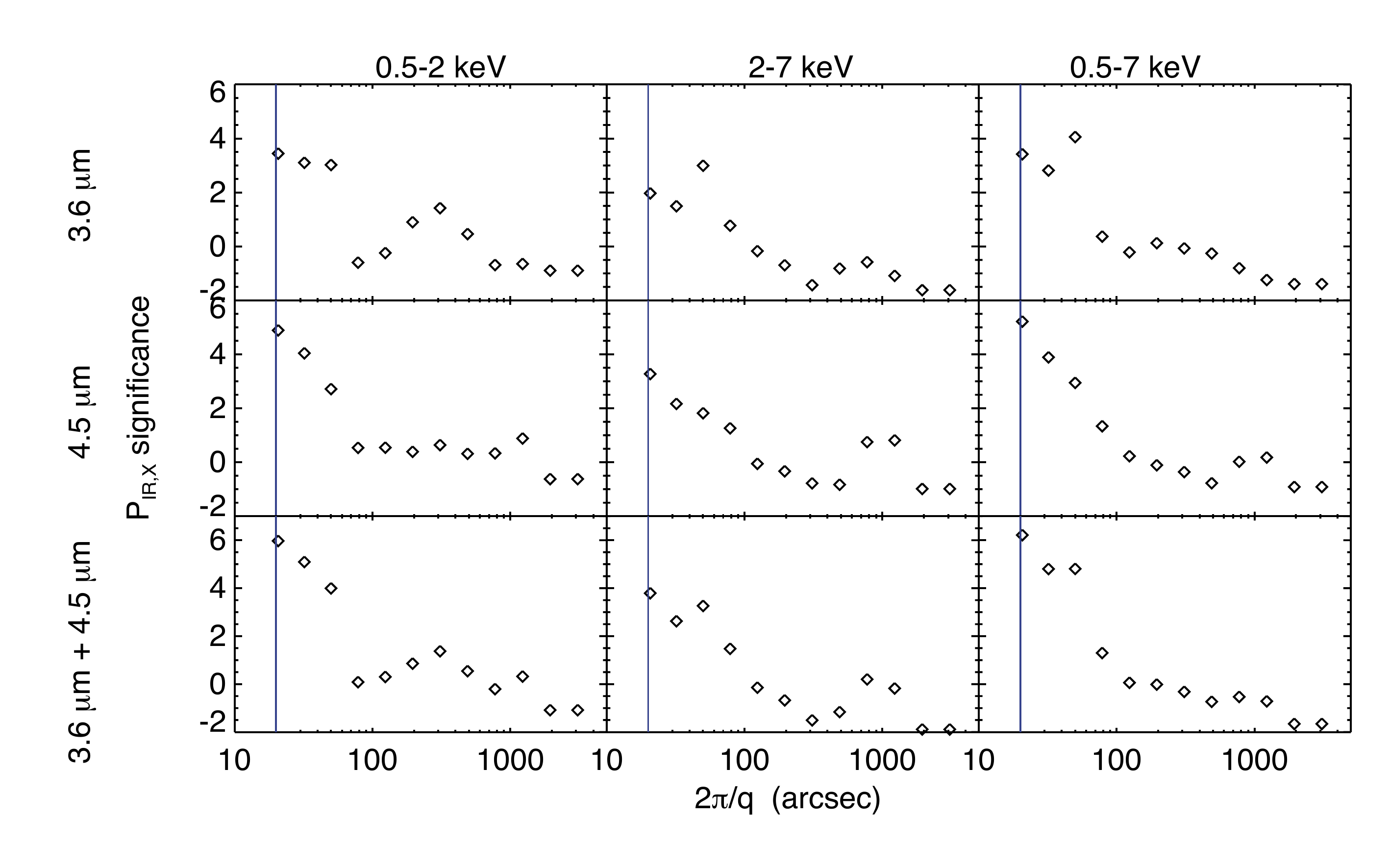}
    \caption{The significance level of the mean cross-power (i.e., $P_{IR,X}$/$\sigma$$_{P_{IR,X}}$) as a function of the angular scale above which the mean cross-power is calculated. In this study, the mean power is calculated above 20$''$ (blue line), same as previous studies \citep[][and C17]{Cappelluti13,Mitchell-Wynne16}.}{\label{fig:sig_trend}}
      \end{figure*}
With similar techniques of Fourier analysis described in Section~\ref{s:irfft}, for each IR \& X-ray band, we extract their cross power spectrum, in each east \& west subfield, respectively. The final power spectra for each pair of the IR and X-ray band are the average among the results from the 2 subfields. We also introduce a combined IR band  (3.6 + 4.5) \mum\ and calculate the cross power between the combined IR band  and each X-ray band, which is the weighted average of the cross power spectrum from 3.6 and 4.5 \mum. 

In Figure~\ref{fig:irx}, we present the CIB-CXB cross power spectra between IR wavelengths and soft, hard \& broad X-ray bands. In Figure~\ref{fig:irx_narrow}, we show the results with narrow X-ray bands. 
For comparison with previous results, we focus on the following analysis of the spectra of soft, hard \& broad X-ray bands.

Previous measurements are limited to scales less than $\sim$1000$''$. Although in this study we extend that to $\sim$ 3000$''$, our cross-power measurements are better constrained within $\sim$ 1000$''$ where we have robust statistics and at large scales the powers are mostly negative and not significant. In order to evaluate the overall significance of our cross power, we calculate the weighted mean cross power ($\langle$$P_{IR,X}$$\rangle$) above 20$''$ and the results are listed in Table~\ref{tab:irx}. The soft X-ray has good cross correlation with both IR wavelengths at $\sim$~4$\sigma$ significance level. The hard band shows a marginal correlation with the 3.6\mum\ data, whereas it's more correlated with the 4.5\mum\ at $\sim$ 4$\sigma$ significance level. The cross powers between broad band X-ray data and both IR wavelengths are at $\sim$4-5$\sigma$ levels. On the other hand, the 4.5\mum\ data in general show stronger correlation than the 3.6\mum\ data with all X-ray bands with $\gtrsim$~4$\sigma$~significance level. Looking at the combined IR data and X-ray data confirms the significant cross correlation between the CIB and the CXB at all X-ray bands. 

We also calculate the mean power of the IR signal vs. X-ray noise cross power ($\langle$$P_{IR,A-B}$$\rangle$), which indicates the level of random error in our analysis. Most of the values are systematically lower than the mean power of the signal-signal cross power (shown in the Table~\ref{tab:irx}). 

We find that the significance of the mean cross-power is very sensitive to the scale range over which the power is calculated. In this work, as well as previous studies \citep[][C17]{Cappelluti13,Mitchell-Wynne16}, the mean power is calculated above 20$''$. We find that if the mean power is calculated above a larger angular scale, it is less significant. In Figure~\ref{fig:sig_trend}, we show how the significance varies with the lower limit of the mean power calculation range, starting with 20$''$. It tells us that including the emission from around tens of arcseconds when calculating the mean cross power is important as emissions from larger scales alone would not give us as significant results.

We overlay the measurements from C17 in our Figure~\ref{fig:irx} with blue symbols and their measurements seem mostly consistent with ours. The two data sets are at similar shot-noise levels as well, with ours at m$_{AB}$$\sim$24.2 mag and theirs at 24.8 mag. In order to quantitatively compare our and C17's measurements, we perform Kolmogorov-Smirnov tests and compute the p-values (the probabilities of having the values of D statistic as large or larger than observed) for all measurements above 20$''$. For 3.6\mum\ vs soft \& hard X-ray and 4.5\mum\ vs hard X-ray, the p-values are all above $\sim$~0.7, and for 4.5\mum\ vs soft X-ray, the p-value is $\sim$~0.2. Statistically, all these p-values are so high that we cannot reject the hypothesis that these two sets of measurements are from the same distribution.

The lower significance in our study ($\sim$3-4$\sigma$) comparing to C17 ($\sim$5-6$\sigma$) could be resulted from different depths of our IR maps. In C17, the mean depth of coverage of their IR data is $\sim$10~h/pixel, whereas in our work it is $\sim$ 4.4~h/pixel, which corresponds to a signal-to-noise ratio of about 1.5 times lower. 

In addition, in C17, they reported significant cross power between the CIB (3.6\mum\ and 4.5\mum) and CXB in the soft band, but no significant correlation with the hard X-ray band was reported. 
However, in this work we find more contribution from the hard band to the 4.5\mum\ CIB fluctuations, which was also suggested by \citet{Mitchell-Wynne16}. More discussion about the physical implications of the cross power spectra are presented in Section~\ref{s:discussion}.


  \begin{figure}
\centering
      \includegraphics[angle=0, trim=10 0 200 0, clip, width=0.5\textwidth]{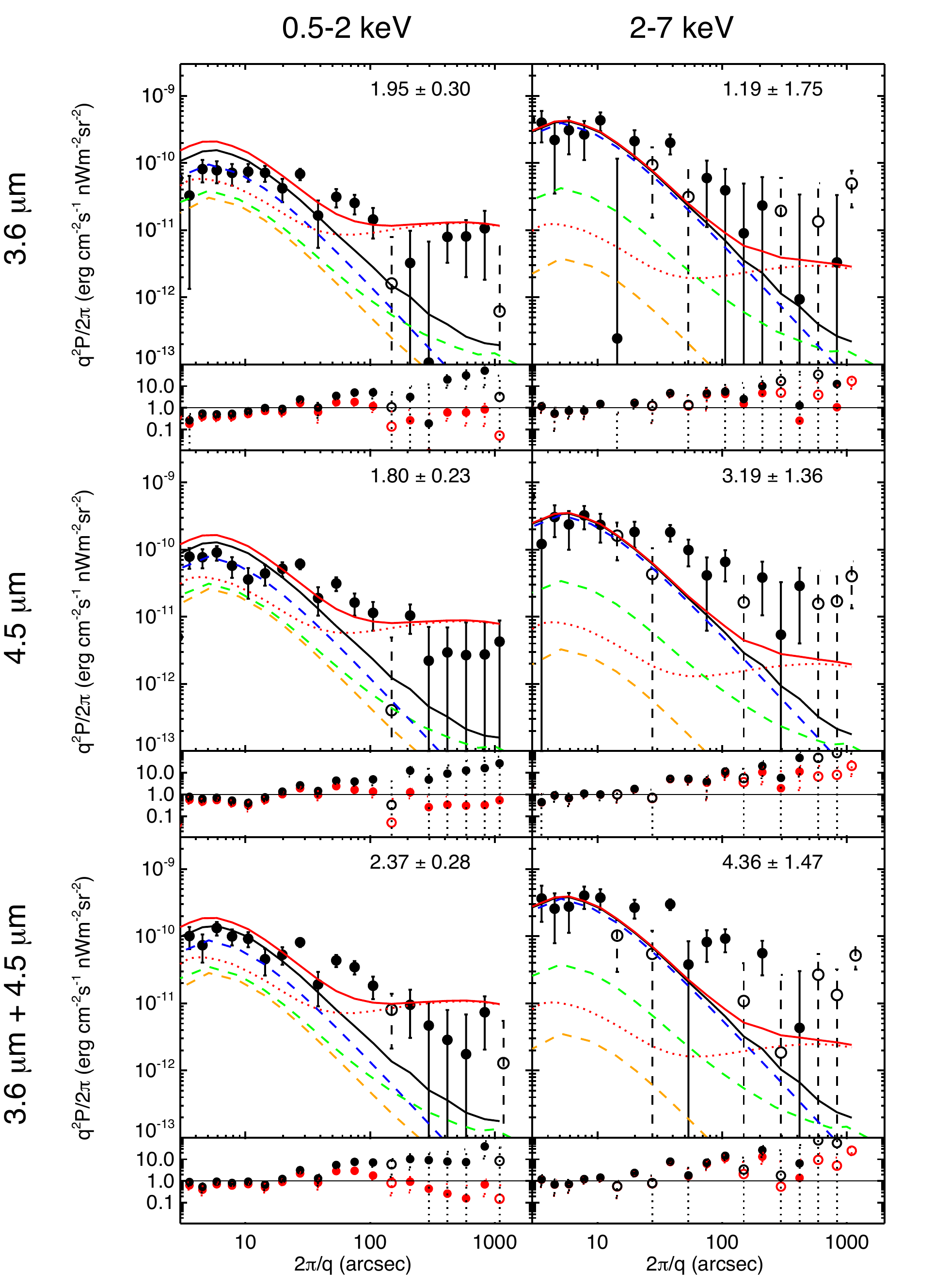}
    \caption{Stacked CIB-CXB cross-power spectra with C17: between 3.6\mum\ and 0.5-2 keV, 2-7 keV (Top panels); between 4.5\mum\ and CXB (Middle panels); combined CIB (3.6\mum\ + 4.5\mum) vs. CXB (Bottom panels). Open circles with dashed error bars denote the absolute values of negative results.
The models are the same as in Figure~\ref{fig:irx}. 
The mean power (calculated above 20$''$) and error are shown in the corner of each panel, in units of 10$^{-19}$~erg~s$^{-1}$~cm$^{-2}$~nW~m$^{-2}$~sr$^{-1}$.}{\label{fig:irx_stack}}
      \end{figure}      
 \subsection{Overall Cross-power After Stacking}\label{s:stacked}
Although our focus of this paper is a thorough study of the cross-power from the COSMOS surveys, we would like to see how much better is the measurement of the clustering component by combining other data. 

As our data are mostly consistent with C17 (Section~\ref{s:cosmos}), we therefore combine the measurements and calculate the stacked cross-power spectra between four photometric band pairs that are available in C17. The stacked spectra are shown in Figure~\ref{fig:irx_stack}. Comparing to Figure~\ref{fig:irx}, the signal-to-noise of the stacked spectra is improved and the excess fluctuations above the predicted values from known (unresolved) populations are more significant. More discussions about the models and the comparison with the data are in Section~\ref{s:model}.



 \section{Discussion}\label{s:discussion}
\subsection{CIB Auto Power}

Theoretically, one might be able to uncover the origin of the CIB excess fluctuations by looking at the CIB auto power spectrum alone, as some models predict different shapes of the spectra at angular scales above $\sim$1000$''$ \citep{Kashlinsky12,Wright98}. 

In this study, although we extend the measurements of the CIB auto power to $\sim$ 3000$''$, the measurements above 1000$''$ are not well constrained and therefore it still appears difficult to constrain various models, e.g., IHL \citep{Cooray12,Zemcov14}.  
However, one could test and differentiate currently constructed IHL models from others by observing the behavior of the clustering component from the models with decreasing shot-noise levels \citep{Kashlinsky15a}. Future studies with deeper IR observations could probe how IHL models handle the data with lower shot-noise levels. A successful model should be able to explain both the large scale fluctuations and the corresponding shot-noise power at small scales.

As in this work we focus on the cross power between CIB and CXB, we do not pursue the detailed analysis of the CIB auto power spectra. 

\subsection{Cross-power: Unresolved Populations}
\label{s:model}
In order to explain the measured CIB-CXB cross power, we discuss some possible source populations, including unresolved AGNs and galaxies, DCBHs \& PBHs (Section~\ref{s:sed}). 

A cross-power signal can arise from a population of sources that emit both in IR and X-rays or share the same environments. Known sources of extragalactic X-rays include i) normal galaxies, ii) AGN and iii) hot gas in clusters and groups. In the following, we use the cross-power reconstruction of \citet{Helgason14} with some improvements.

Galaxies contain high- and low-mass X-ray binaries whose X-ray luminosities have been found to scale with star formation rate and stellar mass respectively \citep[e.g.][]{Basu13}:
\begin{equation} \label{relation}
  L_{\rm X} = \alpha {\rm SFR} (1+z)^\gamma + \beta M_\star (1+z)^\delta ~, 
\end{equation}
where $\alpha$, $\beta$, $\gamma$ and $\delta$ are parameters for which we adopt the values measured by \citet{Lehmer16} in both the soft and hard band, taking into account the scatter in the relation.
For the galaxy population we use a semi-analytic galaxy formation model based on the Millennium simulation \citep{Henriques15}, which reproduces the observed star formation history and stellar mass function as a function of redshift.

We use the IR brightness and a projected position given by the model light cones to create a model image. The brightness distribution is also in a good agreement with observed galaxy counts in the near-IR. We assign an X-ray brightness to the same image position according to Equation \ref{relation} and the luminosity distance of the source. To mimic the source masking, we eliminate all sources with IR magnitude brighter than m$_{AB}$=24.2 and 24.3 at 3.6 and 4.5${\rm \mu m}$ respectively. {\new These magnitude limits are tuned to match the shot noise level in the IR auto power spectrum, which is understood to be galaxy-dominated.} On large scales however, the IR auto power spectrum of the model is lower than measurements and is in agreement with \citet{Helgason12a}.

For the AGN contribution, we adopt the population model of \citet{Gilli07a} in X-rays and \citet{Helgason14} in IR. The extent to which AGN are removed by the joint IR/X-ray mask is estimated using empirical X-ray-to-optical relations (\citealt[for details see]{Civano12} \citealp{Helgason14}). We include the large scatter found in these relations which reflects the widely varying properties of AGNs and their host galaxies.

Hot X-ray emitting gas in groups and clusters of galaxies spatially correlates with IR emitting sources sharing the same environments. We adopt the hot gas modeling of \citet{Helgason14} (Section~5.1.3) which uses the mass and extent of hot gas from the same semi-analytic model used for galaxies above \citep{Guo11, Henriques15}. To calculate the total X-ray flux, we assume a beta-model density profile and a Bremsstrahlung spectrum determined by the gas temperature. Because the flux depends sensitively on the gas density, which is very uncertain, we rescale the average gas mass in halos to match observed X-ray group/cluster counts in the ECDF-S \citep{Finoguenov15}. To mimic the masking of clusters and groups, we adopt the 50\% detection completeness level in the ECDF-S.

Because the host galaxy emission dominates faint AGN in the IR, we only consider the cross correlation of galaxies with each of the three X-ray source classes, described above. 

In Figure~\ref{fig:irx}, we show the measured cross-power spectra with respect to the models for known (unresolved) populations. We find that existing models (the total of AGNs, galaxies and hot gas) described above are sufficient to account for the measured cross power, given relatively large uncertainties in our data. 

\begin{figure}
\centering
      \includegraphics[trim=0 30 0 0, clip, width=0.5\textwidth]{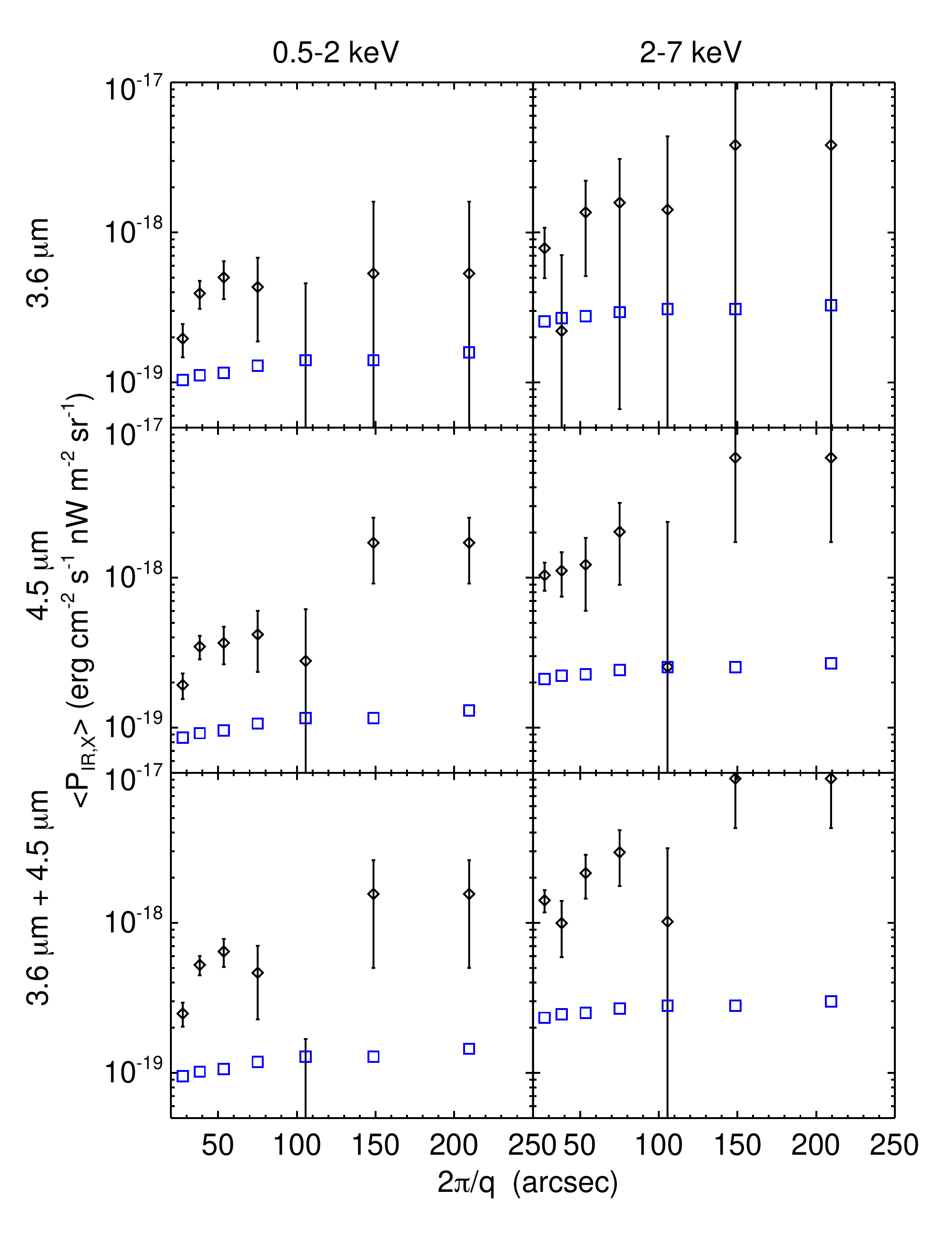}

    \caption{ $\langle$P$_{IR,X}$$\rangle$ as a function of the {\new lower threshold of the angular scale range over which the mean powers are calculated. Each data point is the mean power calculated between the lower threshold and a upper limit of 300$''$. }The black points are calculated from the stacked spectra in Figure~\ref{fig:irx_stack}, while the blue squares are calculated from the models of total known (unresolved) populations. }{\label{fig:stack_trend}}
      \end{figure}

In the stacked spectra (Figure~\ref{fig:irx_stack}), however, the excess above the models of known populations is more significant. To better visualize the discrepancy between the measurements and the model predictions, we calculate the mean cross-power $\langle$P$_{IR,X}$$\rangle$ above different scales (with an upper limit of 300$''$), from both the data and the models for known populations. The results are shown in Figure~\ref{fig:stack_trend}. For clarity, we only show the results within 300$''$, above which the results are too noisy due to our present data quality. 
As shown in the figure, the magnitudes of the excess vary from different pairs of the cross-power and at different angular scales, but overall the models under-estimate the cross-power by one order of magnitude on average, within 300$''$.

\subsection{Cross-power: High-$z$ BHs}\label{s:sed}
{\new Two types of theories about the high-$z$ BHs are possible to explain the observed cross power, as briefly discussed below. }One possibility are the PBHs \citep{Carr75}, which may have formed in the early Universe at the end of the inflationary phase within the mass range of $\sim$~10 $-$ 100~M$_\odot$. If the recently observed massive BH population observed with LIGO \citep{Abbott16b,Abbott16a} is interpreted in terms of PBH, they may be able to explain the measured excess cross power. As proposed by \citet{Kashlinsky16}, if the dark matter is made up of primordial black holes, then it would add small scale power of the density field and increase the accreting efficiency during the collapse of the first halos at high redshift. 

Alternatively, \citet{Yue13} proposed highly obscured DCBHs (M = 10$^{4-6}$ M$_\odot$) at $z$ $>$12 as the origin of the cross power of the CIB and CXB. They may have formed in the first star era in metal-free halos but have not been detected so far by any X-ray deep surveys \citep{Cowie12}. If DCBHs could explain all of the CIB fluctuations at 3.6\mum\ and 4.5\mum, at optical wavelengths there should be little fluctuations caused by these DCBHs as their Lyman break redshifted to the infrared today. Consistent with this expectation, \citet{Mitchell-Wynne16} found that both soft and hard X-ray bands are not significantly correlated with the optical or infrared (0.6 to 1.6\mum) fluctuations. 

    \begin{figure}
\centering
\includegraphics[trim=20 0 0 0, clip, width=0.5\textwidth]{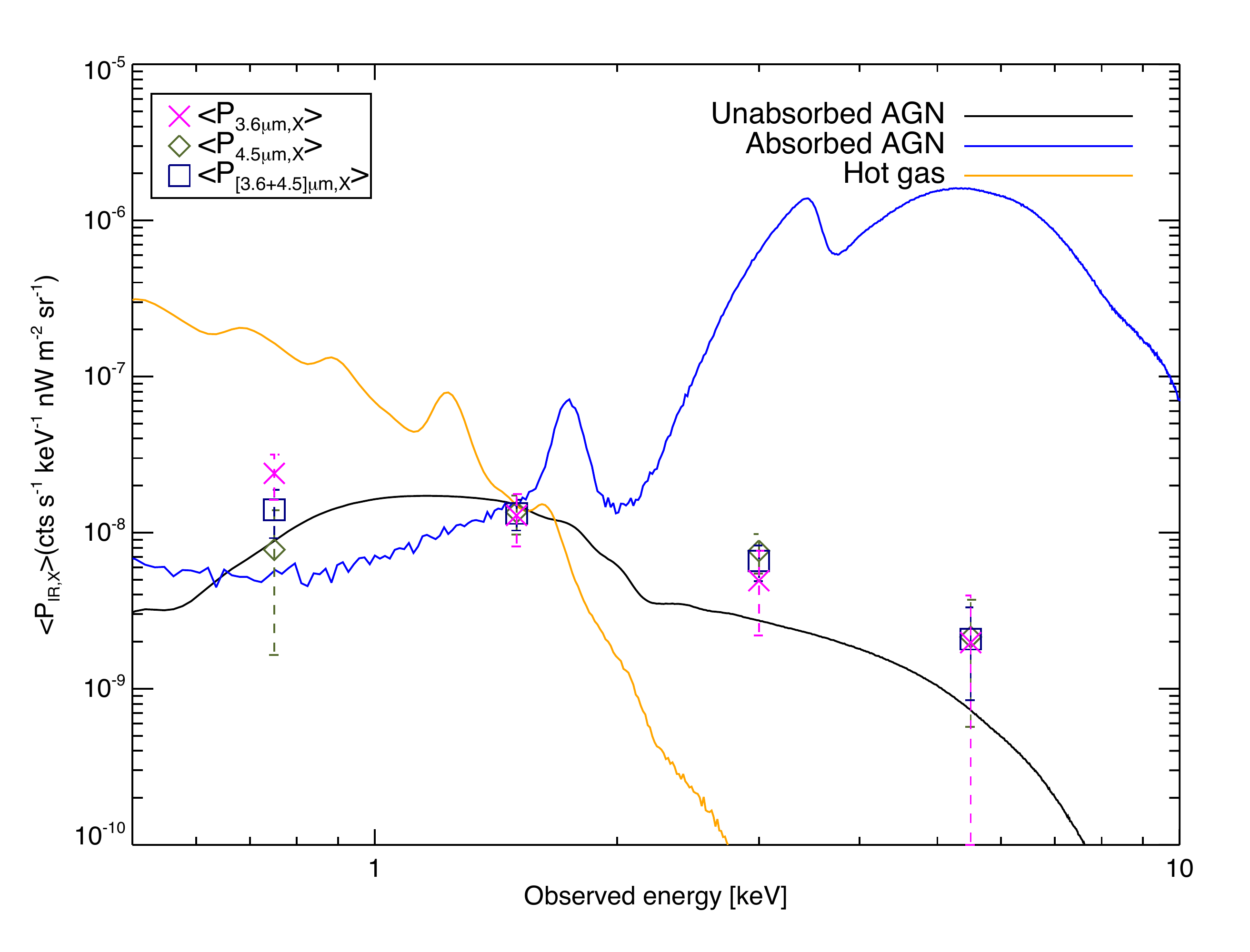}
    \caption{The mean cross power (calculated above 20$''$) between 3.6\mum~(magenta), 4.5\mum~(olive green), and [3.6+4.5]\mum~(blue) and narrow X-ray bands of [0.5-1] keV (0.75 keV), [1-2] keV (1.5 keV), [2-4] keV (3 keV), [4-7] keV (5.5 keV) as a function of the observed X-ray energy. We also show the X-ray spectral models folded through the $\chandra$ response of unabsorbed AGN (black), absorbed AGN (blue) and hot gas (orange). All of the model SEDs are normalized at 1.5 keV. \label{fig:irx_sed_xspec}}
      \end{figure}
    \begin{figure}
\centering
                  \includegraphics[trim=0 0 8 0, clip, width=0.53\textwidth]{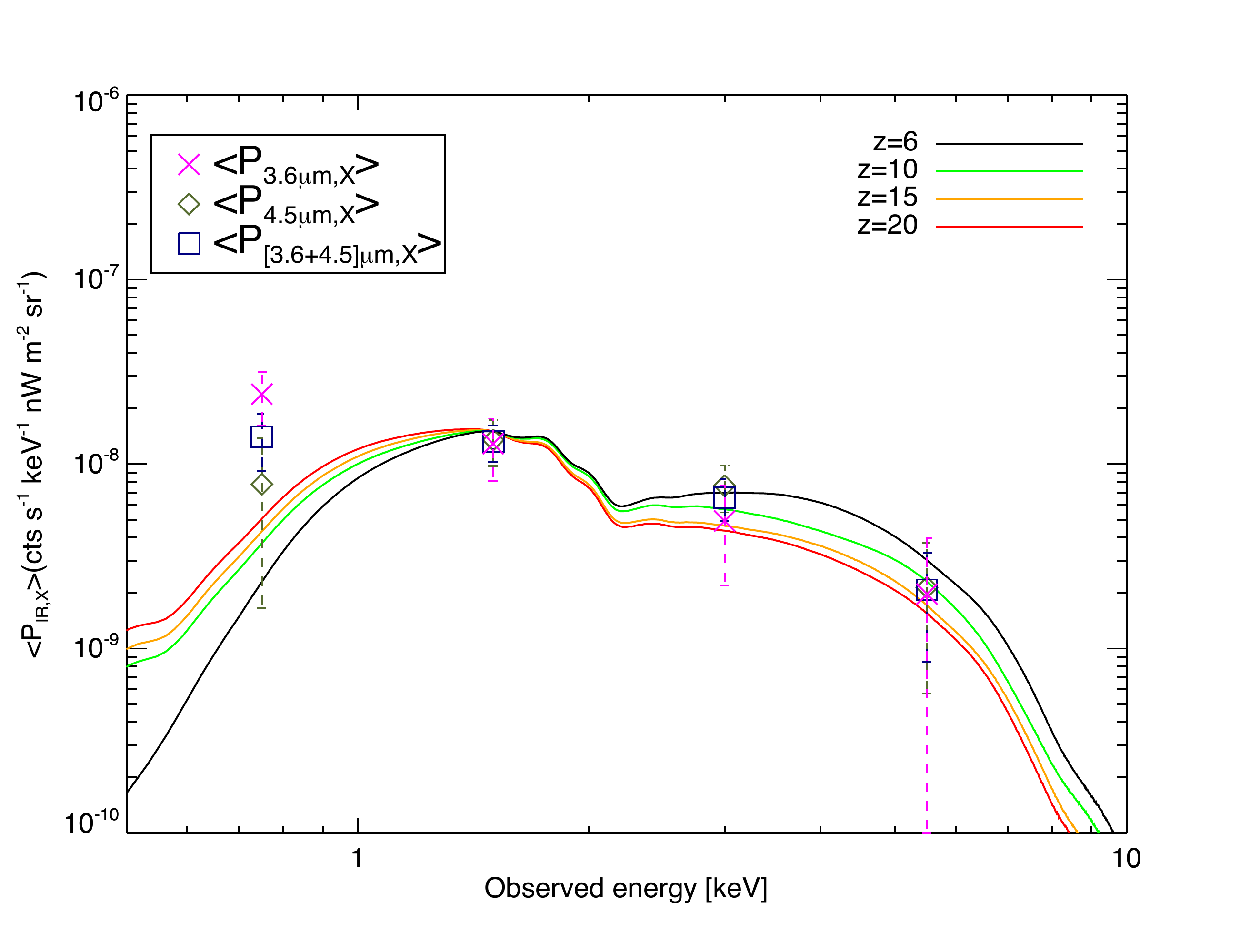}
    \caption{Same data as in Figure~\ref{fig:irx_sed_xspec}, with the DCBH models (at $z$ = 6, 10, 15, 20) folded through the $\chandra$ response and normalized at 1.5 keV, in black, green, orange and red, respectively. \label{fig:irx_sed_dcbh}}
      \end{figure}

As shown in Figure~\ref{fig:irx} and \ref{fig:irx_stack}, we find that adding in the signal from DCBHs, the resulted model spectra are still consistent with our measurements at large scales, especially at the soft X-ray band. In the hard X-ray band, however, it seems that current DCBH models under-estimate the cross power. 

Another approach to diagnose the cross-power signal is to study the X-ray spectral shape of the cross-power signal, under the assumption that the coherence between IR and X-ray is the same for every X-ray band. Separating the CXB into narrower bands and calculating the cross power for each of them with the CIB fluctuations, could then allow a coarse determination of the X-ray sources. Different models about the origin of the X-ray background emission, e.g., from Pop III stellar X-ray binaries or from DCBH accreting from a Compton-thick gas cloud, can predict significantly different X-ray absorption, which might be discriminated with coarse X-ray colors. 

Therefore, we calculate and plot the mean cross power (above 20$''$) as a function of the observed X-ray energy for each fixed IR wavelength in Figure~\ref{fig:irx_sed_xspec} and Figure~\ref{fig:irx_sed_dcbh}. For each X-ray band, we have 2 independent measurements at 3.6~\mum\ and 4.5~\mum\ and the combined one.

For comparison, we also present X-ray spectral models of absorbed \& unabsorbed AGNs, hot gas, and DCBHs \citep{Pacucci15}.  All of these models are folded through the $\chandra$ ACIS response function, using {\sc Xspec} version 12.10.0 \citep{Arnaud96} and normalized at 1.5 keV. 

For AGNs, we assume a simple standard power-law model with two absorption components (wabs*zwabs*zpo in {\sc Xspec} notation) at redshift of $z$~=~1. The first component models the Galactic absorption with a fixed Galactic hydrogen column density (N$_H$) of 1.72$\times$10$^{20}$ $cm^{-2}$ for the COSMOS field. The second component represents the AGN intrinsic absorption. We choose common values for the power-law photon index ($\Gamma$~=~1.9) and N$_H$ (N$_H$~=~1.5$\times$10$^{24}$ $cm^{-2}$ for absorbed AGNs, N$_H$=10$^{21}$$cm^{-2}$ for unabsorbed ones) \citep{Caccianiga04}. For the hot gas (Bremsstrahlung spectrum), we use {\sc Xspec} model (wabs*apec) with $kT$ = 0.5~keV, at $z$~=~0.5.  

Overall, we find that the DCBH models can fit the spectrum well but we cannot {\new determine the exact redshift due to our data quality and the similarities of the models of DCBH and some common low-z scenarios.} Currently it's difficult to use the cross-power SED as a distinguishing diagnostic. In the future, with wide and deep surveys in the X-rays (e.g., Athena) and IR (e.g., Euclid) this study can be much improved. 

\section{Summary}\label{s:summary}
In this study, we made the CIB and CXB fluctuation maps with the latest\spitzer SPLASH and $\chandra$ COSMOS Legacy data and measured the cross power between the CIB (3.6 and 4.5\mum) and CXB (7 bands). 
We note that the shallow depth of the IR maps affect our cross-power measurements on the very large scales of $\sim$~1~deg.

{\new We found the statistical evidence for the cross-power }between CIB and soft band CXB (above 20$''$). The cross power with hard X-ray is less significant than the soft X-ray at $\sim$~3$\sigma$ level. The measured cross power could be accounted for by models of a sum of unresolved AGNs, galaxies and hot gas, but we could not rule out other possibilities such as partial contribution from DCBHs and PBHs. The stacked spectra reveal the excess fluctuations above contributions from known populations with better signal-to-noise. We perform a coarse study of the SED of the CIB-CXB cross power and the measurements seem to be consistent with the DCBH models. However, with current data quality and techniques, we are not able to study this in more detail.

To improve such a study, further improvements in the sensitivity especially at even larger scales are required in both the infrared and X-ray. Accordingly, deeper and wider surveys in the X-ray and near IR ($\sim$ 1$-$4 \mum) will be one solution. Other techniques, e.g., Lyman tomography\citep{Kashlinsky15b}, could pin down the exact redshift of the source populations. The NASA approved project LIBRAE\footnote{\href{http://www.euclid.caltech.edu/page/Kashlinsky\%20Team}{http://www.euclid.caltech.edu/page/Kashlinsky\%20Team}} (Looking at Infrared Background Radiation Anisotropies with Euclid) will probe the CIB exploiting the Euclid imaging of the Wide and Deep Surveys at near-IR and visible wavelengths with unprecedented precision and scope. Other forthcoming missions like eROSITA and Athena (for X-rays), WFIRST, JWST (for optical to near-infrared) will also offer more direct observational information and enable new methods to address the nature of the CIB excess fluctuations and the X-ray counterparts.

\vspace*{0.2cm} 
\acknowledgments
Y.L. thanks Len Cowie for useful comments and discussions pertaining to this manuscript. N.C. acknowledges $\chandra$ SAO grant AR6-17017B and AR4-15015B. K.H. acknowledges support from the Icelandic Research Fund, grant number 173728-051. Support for this work was provided in part by NASA through ADAP grant NNX16AF29G. NASA's support for the Euclid LIBRAE project NNN12AA01C is gratefully acknowledged. 

{\it Facilities:} CXO (ACIS), $\spitzer$ (IRAC)

\bibliography{./Papers}
\end{document}